\documentclass[referee]{raa}

\usepackage{amssymb}
\usepackage{lipsum}
\usepackage{gensymb}
\usepackage{listings}
\usepackage{verbatim}
\usepackage{microtype,siunitx,booktabs}
\usepackage{makecell}
\usepackage{array}
\usepackage{newtxtext,newtxmath}
\usepackage[T1]{fontenc}
\usepackage{amsmath}
\usepackage{natbib}
\usepackage{rotating}
\usepackage{multirow}
\usepackage{graphicx,times}
\usepackage{hyperref}
\usepackage{float}

\bibpunct{(}{)}{;}{a}{}{,}

\begin{document}

\title{Gaia DR3 Analysis of Four Open Clusters Toward the Galactic Anticenter}

\volnopage{ {\bf 20XX} Vol.\ {\bf X} No. {\bf XX}, 000--000}
\setcounter{page}{1}

\author{
Sel\c{c}uk Bilir\inst{1}
\and Seval Ta\c{s}demir\inst{2} 
\and Ege Eraydın\inst{3}
\and Deniz Cennet \c{C}ınar\inst{2} 
\and Jeison Alfonso\inst{4}
\and Remziye Canbay\inst{5} 
}

\institute{
Istanbul University, Faculty of Science, Department of Astronomy and Space Sciences, 34119, Beyazıt, Istanbul, Türkiye; {\it sbilir@istanbul.edu.tr}\\
\and
Istanbul University, Institute of Graduate Studies in Science, Programme of Astronomy and Space Sciences, 34116, Istanbul, Türkiye\\
\and
Yeditepe University, Faculty of Science, Department of Physics, 34755, Ata\c{s}ehir, Istanbul, Türkiye\\
\and
Universidad de los Andes, Departamento de Física, Cra. 1 No. 18A-10, Bloque Ip, A.A. 4976 Bogotá, Colombia\\
\and
Akdeniz University, Faculty of Science, Department of Space Sciences and Technologies, 07058, Antalya, T\"{u}rkiye
}

\vs \no
 
\abstract{This study provides a detailed examination of the structural, astrophysical, kinematic, and dynamical properties of the open clusters COIN-Gaia~24, Czernik~24, FSR~0893, and UBC~74, which are located in the opposite direction to the Galactic center. Astrometric, photometric, and spectroscopic data from the {\it Gaia} Data Release 3 catalog were used to ensure a precise characterization of cluster members and their physical properties. Membership determination was performed using the \texttt{UPMASK} algorithm applied to a five-dimensional parameter space, yielding 116, 179, 238, and 387 likely members for each cluster, respectively. Structural parameters were derived by fitting King profiles to the radial density distributions of high-probability members. Astrophysical parameters were estimated through Bayesian Markov chain Monte Carlo isochrone fitting based on \texttt{PARSEC} evolution models, complemented by spectral energy distribution analysis using \texttt{ARIADNE}. The resulting extinctions, distances, metallicities, and ages indicate that these are moderately reddened, intermediate-age clusters located between 1 and 3.5 kpc from the Sun. Mean radial velocities combined with Galactic orbital integrations computed with \texttt{galpy}, show that all four clusters follow nearly circular, low-eccentricity orbits typical of thin-disc populations. All four OCs have dynamical relaxation times of 22-98~Myr. Yet, their ages surpass these times by several factors, particularly in Czernik~24, revealing that they are dynamically evolved systems even though the calculated $T_{\rm E}$ values represent lower limits. The results confirm that these OCs serve as reliable tracers of the Galactic thin disc’s chemical and dynamical evolution.}

\authorrunning{Bilir et al.} 

\titlerunning{Four Open Clusters Toward the Galactic Anticenter}  

\maketitle


\section{Introduction}
\label{sec:introduction}

Open clusters (OCs) are important stellar systems for investigating star formation processes, stellar evolution, and the dynamics and chemical structure of the Galactic disc. As relatively young and metal-rich star clusters, OCs provide reliable metrics for deriving distances, ages, and metallicities that would otherwise be difficult to determine from individual stars \citep{Lada03, Maurya20}. Their common origin from the collapse of a molecular cloud ensures that member stars share kinematic coherence and chemical homogeneity, making them indispensable tracers of the Milky Way's formation and evolutionary history \citep{McKee07, Dias21}.

Recent advances in photometric and astrometric studies, particularly with the {\it Gaia} mission \citep{Gaia16}, have provided unprecedented accuracy in determining the positions, trigonometric parallaxes, and proper motions of OC members. These datasets not only enable precise derivation of astrophysical parameters but also orbital reconstruction and shed light on the interaction between Galactic dynamics, spiral arm structure, and radial migration \citep{Tarricq21, Rangwal24}. Studies have revealed that OC properties vary systematically with Galactic center distances, highlighting the importance of mapping clusters throughout the disc to understand the spatial and temporal evolution of star formation and chemical enrichment \citep{Frinchaboy2013, Reddy20, Netopil22}.

The Galactic anticenter region plays a particularly strategic role in this context. Located opposite the Galactic center, this region provides a clear window into the outer disc, where the influence of the Galactic bar is minimal and large-scale dynamical effects can be investigated with less complexity. This region is important for studying outer disc structure, metallicity gradients, disc flaring and warping, and the transition between thin and thick disc components \citep{Geisler1987, Momany2006, Lepine2011, Monguio2013, Monteiro2020, Hunt2025}. Studies of anticenter clusters have shown that they bear traces of both the chemical evolution of the Galactic disc and its dynamical heating, making them powerful probes of the Galaxy's secular evolution. With the advent of \citet{Gaia23}, systematic analyses of OCs in the anticenter have provided a more detailed picture of the radial metallicity gradient, spiral arm mapping, and kinematic substructures that dominate the outer disc \citep[cf.][]{Gaia-anticenter}.

\begin{figure}
\centering
\includegraphics[width=0.6\linewidth]{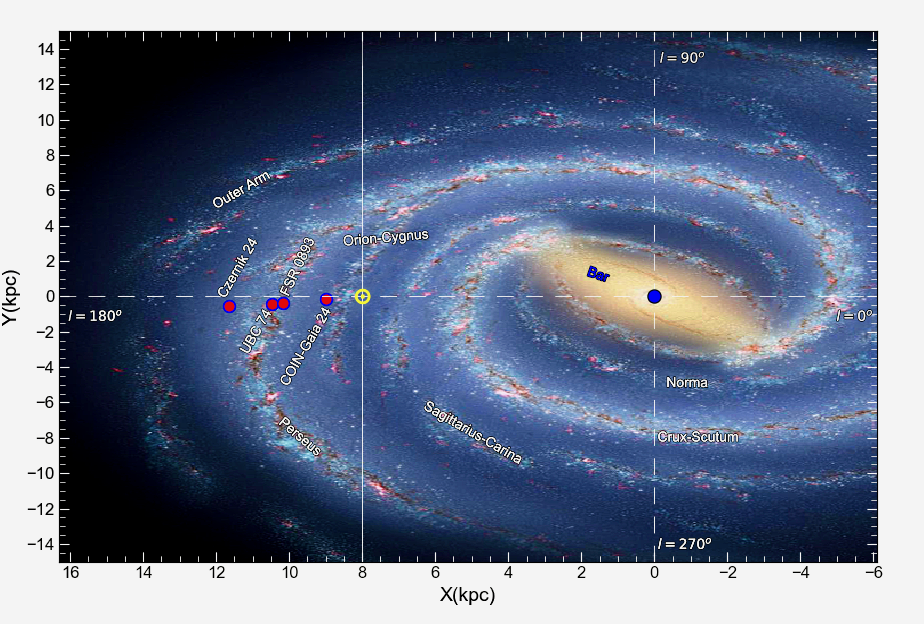}\\
\caption{Projected spatial distribution of the four OCs on the Galactic plane, with the Sun at 8 kpc from the center and each OC shown as a red dot in the Galactic anti-center region.}
\label{fig:map}
\end{figure}

This study presents a detailed analysis of the OCs COIN-Gaia~24, Czernik~24, FSR~0893, and UBC~74, located in the Galactic anticenter direction and only sparsely investigated in previous studies. These clusters were chosen because they occupy similar lines of sight away from the Galactic center while covering a broad interval in heliocentric distance and evolutionary development \citep[e.g.][]{Carraro2003, Carraro2004, Villanova2005}. This combination enables a clearer evaluation of how OC dynamical and chemical characteristics change with Galactocentric radius, with reduced geometric biases. Furthermore, their contrasting dynamical states offer a useful basis for testing the diagnostic power of {\it Gaia} third data release \citep[{\it Gaia} DR3,][]{Gaia23} across varied cluster conditions. The projected positions of the four OCs on the Galactic plane are shown in Figure \ref{fig:map}, followed by a summary of the existing literature.  

{\bf COIN-Gaia~24} ($\alpha = 06^{\rm h} 02^{\rm m} 46^{\rm s}.32$, $\delta = +23^\circ 12' 10''.80$; $l = 186^\circ.89279$, $b = 0^\circ.41579$) was discovered through the analysis of stars in the {\it Gaia} DR2 catalog \citep{Gaia18}. The mean proper-motion components of its 70 member stars were measured as $\mu_{\alpha} \cos \delta = 2.54 \pm 0.10$~mas~yr$^{-1}$ and $\mu_{\delta}=-2.97 \pm 0.09$~mas~yr$^{-1}$, with a trigonometric parallax of $\varpi=0.96\pm 0.05$~mas, corresponding to a heliocentric distance of 1006 pc \citep{Cantat2019}. The first precise determination of the cluster’s fundamental astrophysical parameters was reported by \citet{Monteiro19}, who found an age of $\log t=8.482 \pm 0.298$, a distance of $1149 \pm 82$~pc, a heavy-element abundance of $Z = 0.020 \pm 0.005$, and a colour excess of $E(B-V)=0.392 \pm 0.042$~mag.

{\bf Czernik~24} ($\alpha = 05^{\rm h} 55^{\rm m} 23^{\rm s}.52$, $\delta = +20^\circ 52' 33''.60$; $l = 188^\circ.05847$, $b = -2^\circ.22632$) was discovered by \citet{Czernik66} during his inspection of faint clusters on the Palomar Sky Atlas that were not included in the catalog of \citet{Alter58}. \citet{Ruprecht66} classified the cluster as IV~3~m according to the Trumpler system, indicating that it is a loose system with a clear contrast against the surrounding stellar field and contains a moderate number of stars. The first detailed astrophysical parameters of Czernik~24 were derived by \citet{Koposov2008} using infrared data from the Two Micron All Sky Survey (2MASS; \citealt{Skrutskie2006}). Their analysis yielded a colour excess of $E(B-V)=0.26$ mag and an age of $t \approx 2.5$ Gyr. With the advent of the {\it Gaia} era, \citet{Cantat-Gaudin18} analyzed {\it Gaia} DR2 data \citep{Gaia18}, identifying 121 likely members of the cluster, and determined the mean proper-motion components as $(\langle \mu_{\alpha}\cos\delta, \mu_{\delta}\rangle) = (0.222 \pm 0.027,\,-2.629 \pm 0.028)$ mas~yr$^{-1}$, and the distance as $d = 3449^{+97}_{-88}$ pc. Furthermore, \citet{Soubiran18} derived a radial velocity of $30.10 \pm 2.78$ km~s$^{-1}$ from the spectroscopic data of a cluster member. 

{\bf FSR~0893} ($\alpha = 06^{\rm h} 13^{\rm m} 48^{\rm s}.96$, $\delta = +21^\circ 36' 28''.80$; $l = 189^\circ.52650$, $b = +1^\circ.87372$) is located in the direction opposite to the Galactic center, and its cluster candidacy was confirmed based on 2MASS stellar counts \citep{Camargo2012, Camargo2013}. Using colour-magnitude diagrams (CMDs) constructed from 2MASS photometry, and fitting \texttt{PADOVA} isochrones \citep{Marigo2008}, the authors reported an interstellar extinction of $A_{\rm V}=0.99 \pm 0.06$ mag, a distance of $d=1.1\pm 0.5$ kpc, and an age of $t= 3\pm 1.5$ Gyr. By analyzing {\it Gaia} DR2 photometric and astrometric data \citep{Gaia18}, \citet{Cantat-Gaudin18} identified 155 likely members of the cluster, deriving mean proper-motion components of $(\mu_{\alpha}\cos\delta, \mu_{\delta}) = (-0.355 \pm 0.013,\,-3.215 \pm 0.013)$ mas~yr$^{-1}$ and a heliocentric distance of $d=2355\pm24$ pc. Furthermore, \citet{Soubiran18} determined a radial velocity of $39.17 \pm 0.18$ km~s$^{-1}$ from seven member stars, and combined these with astrometric data to derive the cluster’s space velocity components, showing that it belongs to the thin-disc population of the Galaxy. 

{\bf UBC~74} ($\alpha = 06^{\rm h} 21^{\rm m} 50^{\rm s}.64$, $\delta = +22^\circ 25' 08''.40$; $l = 189^\circ.69013$, $b = +3^\circ.89684$) belongs to the population of newly discovered OCs identified in the University of Barcelona Cluster (UBC) catalog. This catalog was initiated by \citet{Cantat-Gaudin18} using {\it Gaia}~DR2 data \citep{Gaia18}, and was subsequently expanded through a series of studies \citep{Castro-Ginard2019, Castro-Ginard2020, Castro-Ginard2022}, where machine-learning and density-based clustering algorithms were applied to the high-precision astrometric and photometric data from {\it Gaia}. UBC~74 was first reported by \citet{Castro-Ginard2019}, who identified 582 new stellar aggregates based on {\it Gaia}~DR2, providing its mean proper-motion components of $(\langle \mu_{\alpha}\cos\delta, \mu_{\delta}\rangle) = (1.092 \pm 0.106,\,-2.619 \pm 0.133)$ mas~yr$^{-1}$, a trigonometric parallax of $\varpi = 0.354 \pm 0.049$ mas, and a preliminary distance estimate. Furthermore, \citet{Cantat-Gaudin20} reports the cluster’s mean age to be approximately $t=440$ Myr. A summary of more recent studies concerning the four OCs is provided in Table~\ref{tab:literature}.

In this study, we analyzed the OCs COIN-Gaia~24, Czernik~21, FSR~0893, and UBC~74 by determining the membership probabilities of their stars, mean proper-motion components, and distances. We report the fundamental parameters, luminosity and mass functions, the dynamical properties, and the kinematic and Galactic orbital properties of four OCs via {\it Gaia} DR3 \citep{Gaia23}. 

\begin{table*}
\setlength{\tabcolsep}{5pt}
\renewcommand{\arraystretch}{0.8}
\scriptsize
  \caption{Fundamental parameters for selected four OCs obtained in this study and compiled from the literature: Colour excess ($E(B-V$)), distance ($d$), iron abundance ([Fe/H]), age ($t$), proper-motion components ($\langle\mu_{\alpha}\cos\delta\rangle$, $\langle\mu_{\delta}\rangle$), radial velocity ($\langle V_{\rm R}\rangle$), and reference (Ref).}
  \begin{tabular}{ccccccccc}
  \hline
  \hline
    $E(B-V)$  & $d$& [Fe/H] & $t$&  $\langle\mu_{\alpha}\cos\delta\rangle$  &  $\langle\mu_{\delta}\rangle$ & $\langle V_{\rm R}\rangle$  & Ref  & \\
    (mag)                       & (pc)                  & (dex)             & (Myr)             & (mas yr$^{-1}$)       & (mas yr$^{-1}$)   & (km s$^{-1})$     &      \\
    \hline
  \hline
  \multicolumn{8}{c}{COIN-Gaia 24} \\
        \hline
    \hline

$0.392\pm0.042$ & $1149\pm82$   & ---            & $302\pm226$        & ---                & ---                & $2.22\pm0.15$      & 01 & \\
---             & $1046\pm40$   & ---            & $63\pm4$           & $2.538\pm0.243$    & $-2.944\pm0.236$   & ---                & 02 & \\
---             & $1006\pm5$    & ---            & ---                & $2.536\pm0.015$    & $-2.954\pm0.011$   & ---                & 03 & \\
0.216           & 1030          & ---            & 83                 & $2.536\pm0.097$    & $-2.954\pm0.092$   & ---                & 04 & \\
$0.338\pm0.020$ & $941\pm27$    & $0.007\pm0.072$& $245\pm129$        & $2.564\pm0.096$    & $-2.960\pm0.094$   & ---                & 05 & \\
---             & 1030          & ---            & 83                 & $2.515\pm0.063$    & $-2.957\pm0.069$   & ---                & 06 & \\
0.387           & $1013\pm31$   & ---            & 63                 & $2.508\pm0.096$    & $-2.951\pm0.067$   & ---                & 07 & \\
---             & 1030          & $0.057\pm0.008$& ---                & $2.536\pm0.097$    & $-2.954\pm0.092$   & $-9.20\pm1.35$     & 08 & \\
---             & $966\pm44$    & ---            & 83                 & $2.495\pm0.110$    & $-2.948\pm0.075$   & $-3.00\pm0.52$     & 09 & \\
$0.357\pm0.014$ & $953\pm12$    & $0.082\pm0.029$& $178\pm63$         & ---                & ---                & ---                & 10 & \\
$0.253\pm0.056$ & $967\pm3$     & ---            & $100\pm49$         & $2.480\pm0.012$    & $-2.953\pm0.012$   & $-2.94\pm3.17$     & 11 & \\

\hline
    \hline
    \multicolumn{9}{c}{Czernik 24} \\
        \hline
        \hline
        
---            & $3449\pm92$     & ---              & ---             & $0.222\pm0.027$ & $-2.629\pm0.028$ & ---              & 03 & \\
0.526          & 3981            & ---              & 2690            & $0.222\pm0.269$ & $-2.629\pm0.268$ & ---              & 04 & \\
$0.744\pm0.019$& $3623\pm173$    & $-0.297\pm0.079$ & $1342\pm102$    & $0.253\pm0.273$ & $-2.668\pm0.301$ & ---              & 05 & \\
---            & 3981            & ---              & 2690            & $0.253\pm0.099$ & $-2.715\pm0.069$ & ---              & 06 & \\
---            & 3981            & $-0.160\pm0.042$ & ---             & $0.222\pm0.269$ & $-2.629\pm0.268$ & $17.91\pm2.69$   & 08 & \\
---            & $3830\pm1440$   & ---              & 2690            & $0.253\pm0.103$ & $-2.712\pm0.074$ & $20.99\pm0.58$   & 09 & \\
$0.636\pm0.075$& $3723\pm91$     & ---              & $1202\pm513$    & $0.260\pm0.011$ & $-2.721\pm0.009$ & $16.01\pm5.45$   & 11 & \\
---            & $3449\pm92$     & ---              & ---             & $0.222\pm0.027$ & $-2.629\pm0.028$ & ---              & 12 & \\
---            & $3449\pm92$     & ---              & ---             & $0.222\pm0.027$ & $-2.629\pm0.028$ & $30.10\pm2.78$   & 13 & \\
---            & ---             & $-0.110$         & 2690            & ---             & ---              & ---              & 15 & \\
---            & $3449\pm92$     & $-0.238\pm0.068$ & ---             & $0.222\pm0.269$ & $-2.629\pm0.268$ & $18.94\pm7.54$   & 16 & \\
$0.690\pm0.070$& ---             & $-0.220\pm0.280$ & $1778\pm627$    & $0.280\pm0.200$ & $-2.670\pm0.240$ & ---              & 17 & \\
---            & 3980            & $-0.120\pm0.040$ & 2700            & ---             & ---              & $22.10\pm0.50$   & 19 & \\

  \hline
  \hline
    \multicolumn{8}{c}{FSR 0893}\\
        \hline
        \hline
    
---             & $2655\pm169$   & ---             & $1020\pm61$    & $-0.344\pm0.160$ & $-3.200\pm0.149$ & ---               & 02 & \\
---             & $2355\pm24$    & ---             & ---            & $-0.355\pm0.013$ & $-3.215\pm0.013$ & ---               & 03 & \\
0.710           & 2328           & ---             & 562            & $-0.355\pm0.147$ & $-3.215\pm0.150$ & ---               & 04 & \\
$0.856\pm0.064$ & $1888\pm164$   & $-0.097\pm0.097$& $383\pm1025$   & $-0.349\pm0.163$ & $-3.199\pm0.160$ & $39.18\pm0.42$    & 05 & \\
---             & 2328           & ---             & 562            & $-0.322\pm0.067$ & $-3.185\pm0.055$ & ---               & 06 & \\
---             & 2328           & $-0.180\pm0.015$& ---            & $-0.355\pm0.147$ & $-3.215\pm0.150$ & $35.92\pm1.24$    & 08 & \\
---             & $2159\pm233$   & ---             & 562            & $-0.324\pm0.074$ & $-3.185\pm0.071$ & $39.07\pm0.59$    & 09 & \\
$0.941\pm0.034$ & $1957\pm105$   & $0.125\pm0.047$ & $387\pm68$     & ---              & ---              & ---               & 10 & \\
$1.023\pm0.072$ & $2092\pm12$    & ---             & $181\pm93$     & $-0.321\pm0.008$ & $-3.180\pm0.008$ & $37.76\pm1.45$    & 11 & \\
---             & $2355\pm24$    & ---             & ---            & $-0.355\pm0.013$ & $-3.215\pm0.013$ & ---               & 12 & \\
---             & $2355\pm24$    & ---             & ---            & $-0.355\pm0.013$ & $-3.215\pm0.013$ & $39.17\pm0.18$    & 13 & \\
---             & 2191           & ---             & 562            & $-0.355\pm0.013$ & $-3.215\pm0.013$ & $39.17\pm0.18$    & 18 & \\
  \hline
  \hline
    \multicolumn{8}{c}{UBC 74}\\
        \hline
        \hline
        
---            & $2597\pm38$   & ---             & ---            & $1.076\pm0.013$ & $-2.616\pm0.014$ & ---              & 03 & \\
0.445          & 2570          & ---             & 436            & $1.076\pm0.123$ & $-2.616\pm0.128$ & ---              & 04 & \\
---            & 2570          & ---             & 436            & $1.024\pm0.068$ & $-2.600\pm0.067$ & ---              & 06 & \\
---            & 2570          & $-0.098\pm0.013$& ---            & $1.076\pm0.123$ & $-2.616\pm0.128$ & $24.10\pm1.20$   & 08 & \\
---            & $2469\pm299$  & ---             & 436            & $1.025\pm0.075$ & $-2.599\pm0.065$ & $41.57\pm0.75$   & 09 & \\
---            & $2824\pm389$  & ---             & ---            & $1.092\pm0.103$ & $-2.619\pm0.133$ & $43.98\pm1.53$   & 14 & \\
---            & 2496          & ---             & 436            & $1.076\pm0.013$ & $-2.616\pm0.014$ & $42.95\pm0.98$   & 18 & \\
        \hline
        \hline     
    \end{tabular}
    \\
(01) \citet{Monteiro19}, (02) \citet{Liu19}, (03) \citet{Cantat-Gaudin_Anders20}, (04) \citet{Cantat-Gaudin20}, (05) \citet{Dias21}, (06) \citet{Poggio2021},  (07) \citet{He2022}, (08) \citet{Fu22}, (09) \citet{Gaia23}, (10) \citet{Almeida2023}, (11) \citet{Hunt2024}, (12) \citet{Cantat-Gaudin18}, (13) \citet{Soubiran18}, (14) \citet{Castro-Ginard2019}, (15) \citet{Viscasillas2022}, (16) \citet{Zhong20}, (17) \citet{Angelo2021}, (18) \citet{Tarricq21}, (19) \citet{Magrini2021}.
  \label{tab:literature}%
\end{table*}%


\section{Data analysis}
\subsection{Photometric Data}
\begin{figure}
\centering
\includegraphics[width=0.6\linewidth]{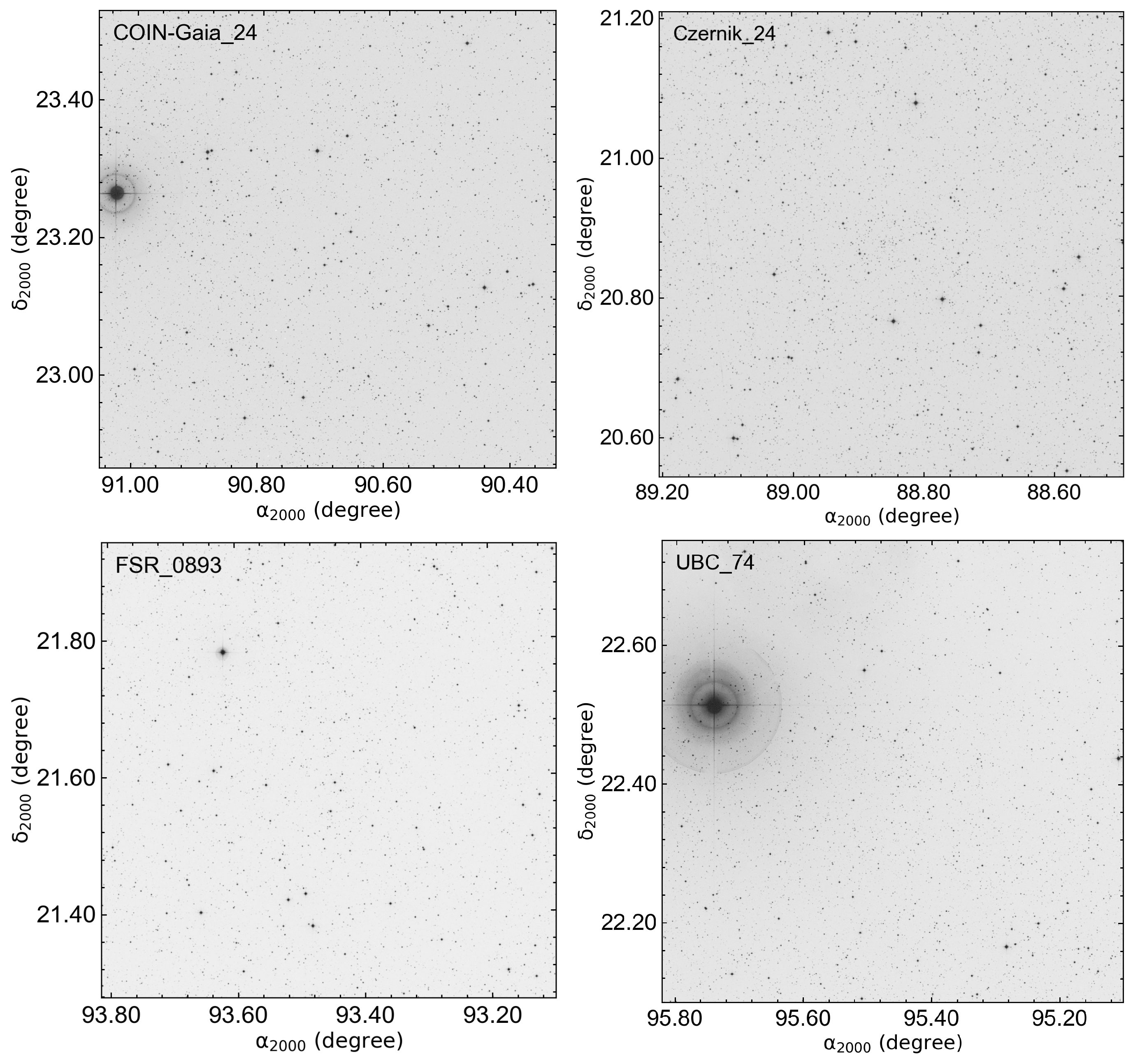}
\caption{$40\times40$ arcmin$^{2}$ star fields for selected OCs in the equatorial coordinate system. North and east correspond to the up and left directions, respectively.} 
\label{fig:charts}
\end{figure}

To visualize the spatial distribution of the studied systems, an identification map was generated. The diagram spans an area of {\bf approximately $40^\prime \times 40^\prime$ centered} on the Galactic anti-center region, clearly delineating the positions of COIN-Gaia~24, Czernik~24, FSR~0893, and UBC~74. The final identification chart is presented in Figure~\ref{fig:charts}, where the four clusters are discernible as distinct stellar concentrations within the observed field. The star charts, obtained using the \texttt{POSS2UKSTU\_RED} filter\footnote{\url{https://archive.stsci.edu/cgi-bin/dss_form}}. The analysis employs astrometric, photometric, and spectroscopic data from the {\it Gaia} DR3 archive \citep{Gaia23}. The photometric dataset comprises the $G$, $G_{\rm BP}$, and $G_{\rm RP}$ passbands, whereas the astrometric dataset includes equatorial coordinates ($\alpha, \delta$), trigonometric parallaxes ($\varpi$), and proper-motion components ($\mu_{\alpha} \cos \delta, \mu_{\delta}$), with radial velocity measurements ($V_{\rm R}$) incorporated if available. All retrieved parameters were accompanied by their respective uncertainties and treated consistently throughout the study. From circular regions with a radius of 40 arcminutes centered on each cluster, the numbers of stars retrieved from the {\it Gaia} DR3 database were 57,089 for COIN-Gaia~24, 61,394 for Czernik~24, 46,970 for FSR~0893, and 49,135 for UBC~74.

\subsection{Determination of the Cluster Center}
\label{sec:Cluster_Center}
\begin{figure}
    \centering
    \includegraphics[width=0.7\linewidth]{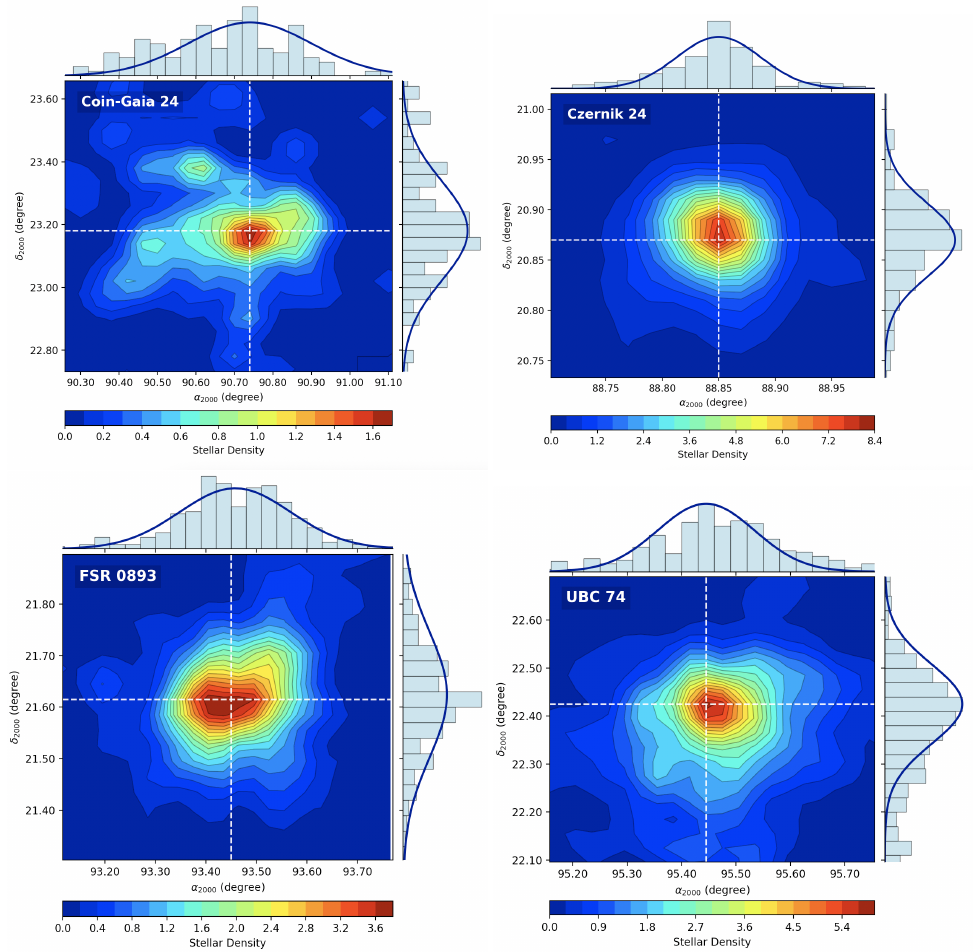}
    \caption{The two-dimensional stellar density maps of the four OCs. The top panel of the density maps shows the histogram for right ascension ($\alpha$), and the right panel shows the histogram for declination ($\delta$). The solid lines represent the Gaussian curves fitted to the histograms.}
    \label{fig:centers}
\end{figure}

The central region of a stellar cluster corresponds to the location with the highest stellar density when the equatorial coordinates ($\alpha, \delta$) of all stars, extracted from the {\it Gaia} archive, are plotted. To identify this position, one-dimensional histograms of $\alpha$ and $\delta$ were generated for each OC. The number of stars in each interval was counted, and Gaussian functions were fitted to the resulting distributions. The peaks of these Gaussian fits in both $\alpha$ and $\delta$ were adopted as the refined coordinates of the cluster centers. These locations, shown in Figure~\ref{fig:centers}, represent the points of maximum stellar concentration for the respective clusters. The equatorial coordinates calculated for the OCs and their Galactic coordinates are given in Table~\ref{tab:Final_table}.

\subsection{Photometric Completeness Limit}
\label{sec:Completeness}
A reliable determination of the structural and astrophysical properties of the four OCs requires first establishing the photometric completeness threshold for each OC. This step is crucial to prevent selection effects that could bias subsequent analyses. In particular, an accurate completeness limit ensures that derived luminosity and mass functions are not artificially flattened at the faint end, which would directly affect the inferred initial mass function (IMF) slopes and stellar population parameters \citep{Salpeter55}, since low-mass main-sequence stars near the detection limit have a significant effect on isochrone agreement \citep{Moraux2003}.

\begin{figure*}
\centering
\includegraphics[width=0.7\textwidth]{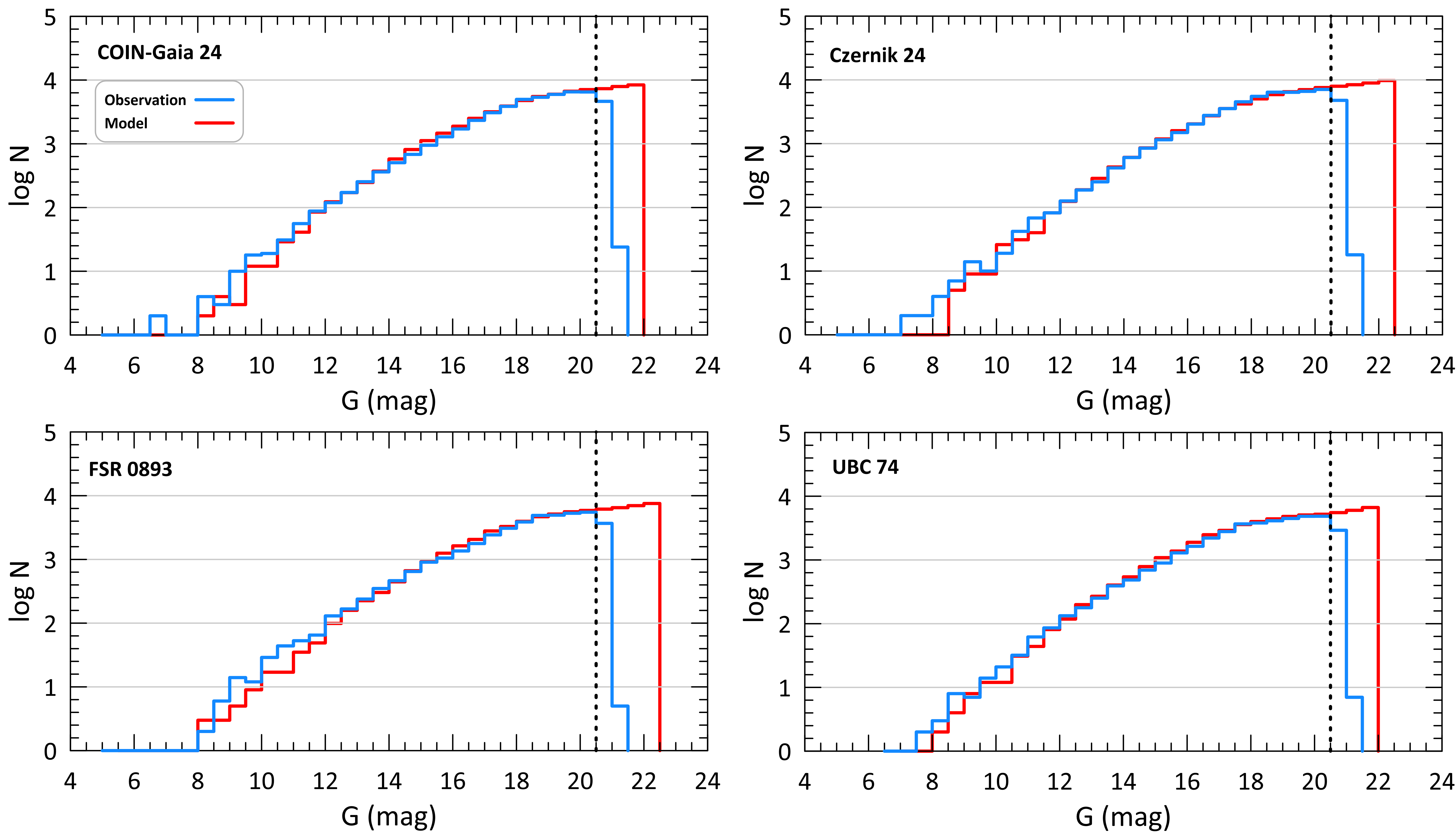}
\caption{Observed (blue) and model (red) stellar count histograms in the $G$-band for the four OCs. The black dashed lines indicate the photometric completeness limits.} 
\label{fig:completness_limit}
\end{figure*}

Stellar number counts were analyzed in successive intervals of the {\it Gaia} $G$-band apparent magnitude. As shown in Figure~\ref{fig:completness_limit}, the stars' distribution increases with magnitude until reaching a maximum near $G=20.5$ mag. Beyond this point, the counts decline, indicating the onset of incompleteness in the data. Consequently, stars fainter than $G=20.5$ mag were excluded from the working sample to ensure reliable parameter estimation for the four OCs.

To evaluate the photometric completeness limits, synthetic stellar populations were created for the areas encompassing both OCs using the Besançon Galaxy Model\footnote{\url{https://model.obs-besancon.fr/}}
 \citep{Robin03}. This model integrates the three-dimensional interstellar extinction map of \citet{Marshall06}, enabling realistic modeling of extinction along various lines of sight. The simulations covered the magnitude interval $5 < G~{\rm (mag)} \leq 23$, matching the distribution of the observed star counts while incorporating the three-dimensional extinction values reported by \citet{Marshall06}. The Besançon Galaxy model calculations were carried out using solid angles of 0.445 square degrees for each OC region.

The synthetic star distributions were then compared with the observational data, as shown in Figure~\ref{fig:completness_limit}. In the brighter part of the $G$ band, where the observed number of stars exceeds that predicted by the model, the main magnitude intervals of the OCs are identified, indicating that completeness is approximately 100\%. In contrast, the apparent magnitudes at which the model begins to overpredict the star counts define the photometric completeness thresholds. Based on this comparison, we confirm that the previously established photometric completeness limits do not exhibit any loss of stars in the present study.

Photometric uncertainties were assessed using the error estimates reported in the {\it Gaia}~DR3 catalog, treated as representative of interval-based uncertainties. Within successive $G$-band bins, mean apparent magnitudes and the mean $G_{\rm BP}-G_{\rm RP}$ colour indices of stars within the OC fields were analysed. At the adopted completeness limit of $G=20.5$ mag, the mean internal uncertainty in the $G$-band was found to be 0.007 mag, while the corresponding mean error in $G_{\rm BP}-G_{\rm RP}$ was calculated to be 0.110 mag. A summary of the mean photometric errors across the full $G$-apparent magnitude interval for the selected OCs is listed in Table~\ref{tab:photometric_errors}. As can be seen from the results in Table~\ref{tab:photometric_errors}, the mean $G$-apparent magnitude and $G_{\rm BP}-G_{\rm RP}$ colour index errors increase as the magnitude decreases. In the analyses conducted up to the limit magnitude of the four OCs, the mean apparent magnitude and colour errors were obtained at almost similar values.

\begin{table*}
\centering
\caption{Errors in $G$-apparent magnitudes and ${G_{\rm BP}-G_{\rm RP}}$ colour indices of stars located along the direction of the selected four OCs. The last row of the table presents the median errors of the $G$-apparent magnitude and colour index for stars within the star counts and completeness limits.}
\label{tab:photometric_errors}
\setlength{\tabcolsep}{2pt} 
\begin{tabular}{c|ccc|ccc|ccc|ccc}
\toprule
 & \multicolumn{3}{c|}{\textbf{COIN-Gaia 24}} & \multicolumn{3}{c|}{\textbf{Czernik 24}} & \multicolumn{3}{c|}{\textbf{FSR 0893}} & \multicolumn{3}{c}{\textbf{UBC 74}} \\
\cmidrule(lr){2-4} \cmidrule(lr){5-7} \cmidrule(lr){8-10} \cmidrule(lr){11-13} $G$ (mag) & $N$ & $\sigma_{\rm G}$ & $\sigma_{G_{\rm BP}-G_{\rm RP}}$ & $N$ & $\sigma_{\rm G}$ & $\sigma_{G_{\rm BP}-G_{\rm RP}}$ & $N$ & $\sigma_{\rm G}$ & $\sigma_{G_{\rm BP}-G_{\rm RP}}$ & $N$ & $\sigma_{\rm G}$ & $\sigma_{G_{\rm BP}-G_{\rm RP}}$ \\
\hline \hline
    6--14  & 944   & 0.003 & 0.006 & 1000  & 0.003 & 0.006 & 931   & 0.003 & 0.007& 992   & 0.003 & 0.007 \\ 
    14--15 & 1015  & 0.003 & 0.006 & 1215  & 0.003 & 0.006 & 928   & 0.003 & 0.006& 1009  & 0.003 & 0.006 \\
    15--16 & 1922  & 0.003 & 0.007 & 2325  & 0.003 & 0.007 & 1771  & 0.003 & 0.008& 1871  & 0.003 & 0.008 \\
    16--17 & 3485  & 0.003 & 0.012 & 4157  & 0.003 & 0.012 & 2719  & 0.003 & 0.013& 3362  & 0.003 & 0.013 \\
    17--18 & 6131  & 0.003 & 0.026 & 7090  & 0.003 & 0.023 & 4914  & 0.003 & 0.029& 5709  & 0.003 & 0.026 \\
    18--19 & 9989  & 0.004 & 0.060 & 11064 & 0.004 & 0.054 & 7720  & 0.004 & 0.069& 8749  & 0.004 & 0.062 \\
    19--20 & 14677 & 0.006 & 0.130 & 15201 & 0.006 & 0.122 & 12168 & 0.006 & 0.144& 12363 & 0.006 & 0.135 \\
    20--21 & 18309 & 0.013 & 0.278 & 18721 & 0.013 & 0.266 & 15442 & 0.013 & 0.287& 14864 & 0.013 & 0.284 \\
    21--23 & 617   & 0.029 & 0.455 & 621   & 0.029 & 0.463 & 377   & 0.028 & 0.490 & 216   & 0.029 & 0.473 \\
    \hline
Total/Error & 57089   &0.007  &0.108  & 61394   & 0.007 & 0.106 & 46970   & 0.007 &0.116 & 49135   & 0.007 & 0.112  \\
    \hline  
\bottomrule
\end{tabular}
\end{table*}

\subsection{Membership Determination}
\label{section:cmds}

The unsupervised photometric membership assignment in stellar clusters \citep[\texttt{UPMASK},][]{Krone-Martins14} algorithm is a model-independent method designed to identify cluster members by statistically distinguishing spatial and kinematic groupings from the field star population. It combines $k$-means clustering with significance tests in multi-dimensional parameter space, iteratively refining membership probabilities without assuming any specific cluster profile. Using \texttt{UPMASK}, the membership probabilities for stars in the fields of the selected four OCs were computed. For each OC, a five-dimensional parameter space was defined using equatorial coordinates ($\alpha, \delta$), trigonometric parallaxes ($\varpi$), proper-motion components ($\mu_{\alpha}\cos\delta$, $\mu_\delta$), and their associated uncertainties as inputs. The membership analysis was conducted iteratively, performing 25 outer-loop iterations per cluster to ensure robust convergence and selecting optimal $k$-means values based on configurations that maximized the distinction between OC members and field stars. Following this procedure, the final $k$-means values were determined as 24 for COIN-Gaia~24, 29 for Czernik~24, 21 for FSR~0893, and 12 for UBC~74.     

\begin{figure}
\centering
\includegraphics[width=0.4\textwidth]{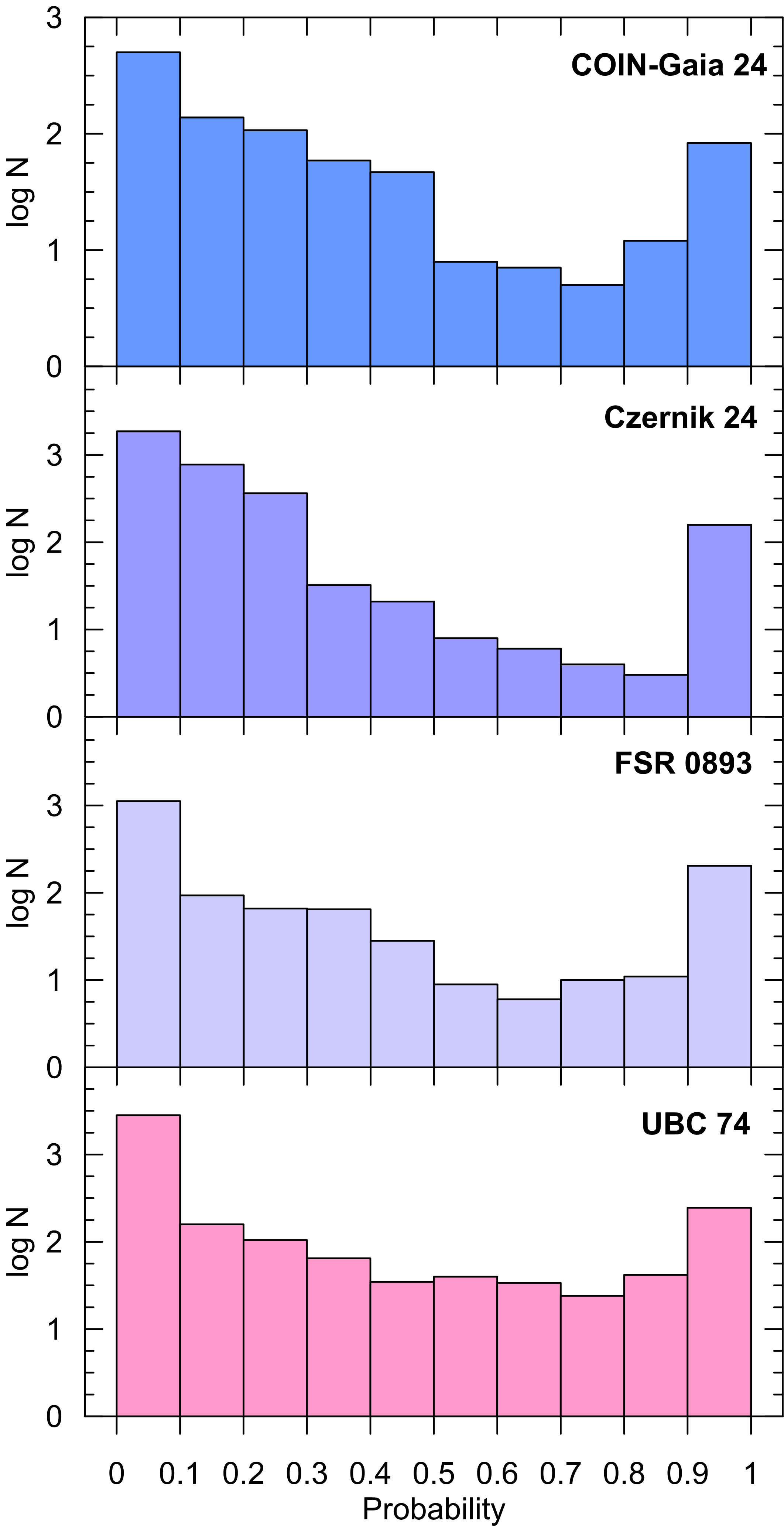}
\caption{The cluster membership probability of stars in the four OC directions.}
\label{fig:P-histograms} 
 \end {figure}

After applying the \texttt{UPMASK} algorithm, stars with membership probabilities ($P$) in the range $0< P\leq 1$ were retained as potential physical members of the clusters. This probabilistic cutoff was selected to maintain an optimal balance between completeness and reliability. The distributions of membership probabilities for four OCs are shown in Figure~\ref{fig:P-histograms}, highlighting a noticeable excess of high-probability stars compared to the background field stars. Following selection based on photometric completeness (with a limiting magnitude of $G\leq 20.5$), 116 stars in COIN-Gaia 24, 179 stars in Czernik 24, 238 stars in FSR 0893, and 387 stars in UBC 74 were identified as most probable cluster members. As shown in Figure~\ref{fig:P-histograms}, most stars exhibit membership probabilities greater than $P=0.5$, indicating a strong likelihood of physical cluster membership. In the subsequent methodology of this study, only stars with membership probabilities of $P\geq 0.5$ were considered.

\begin{figure*}
\centering
\includegraphics[width=1\textwidth]{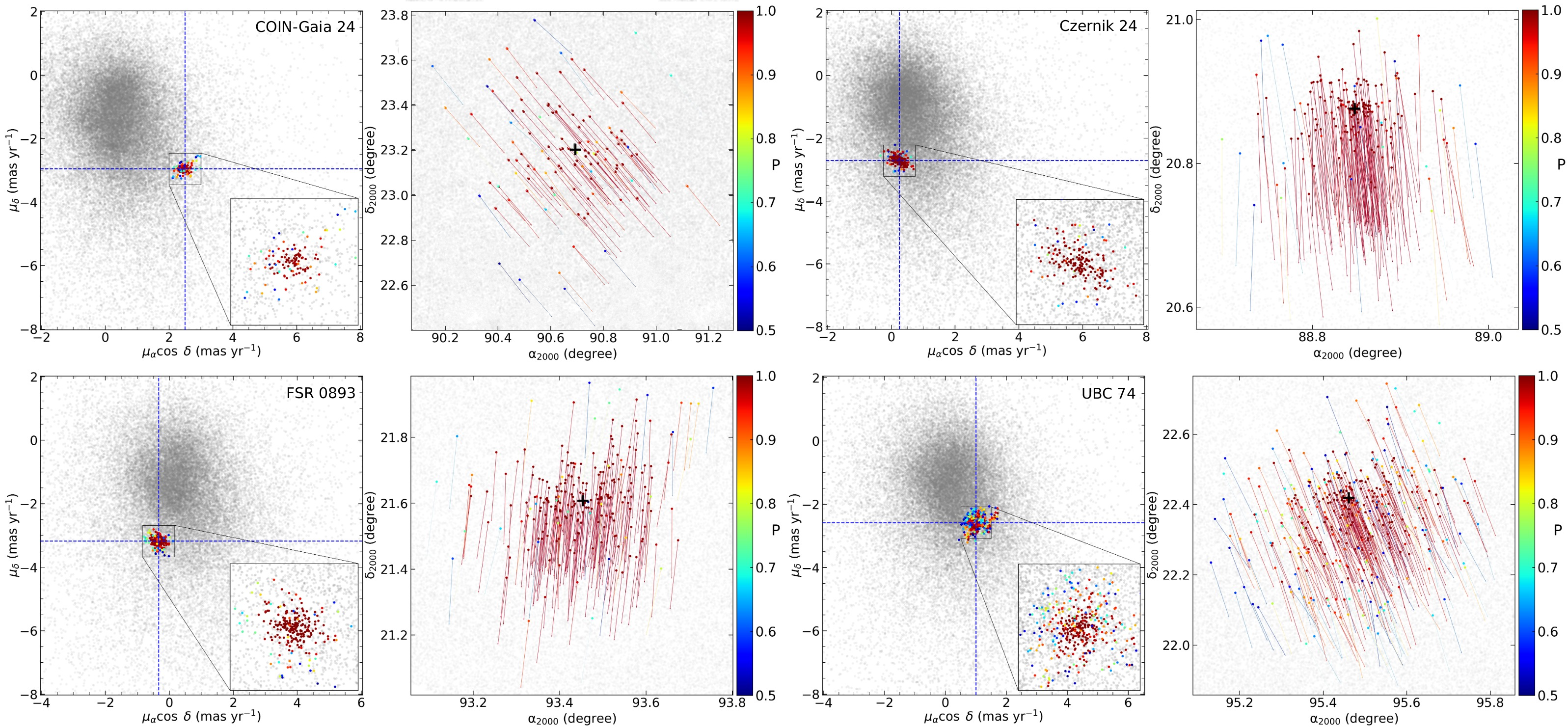}
\caption{Stellar positions in the four OCs on the VPDs, along with their proper-motion vectors projected onto the equatorial coordinate planes. While the black plus signs show the central equatorial coordinates of the OCs, blue dashed lines represent the median proper-motion components.}
\label{fig:VPD_all} 
 \end {figure*}

\begin{figure} 
\centering
\includegraphics[width=0.7\columnwidth]{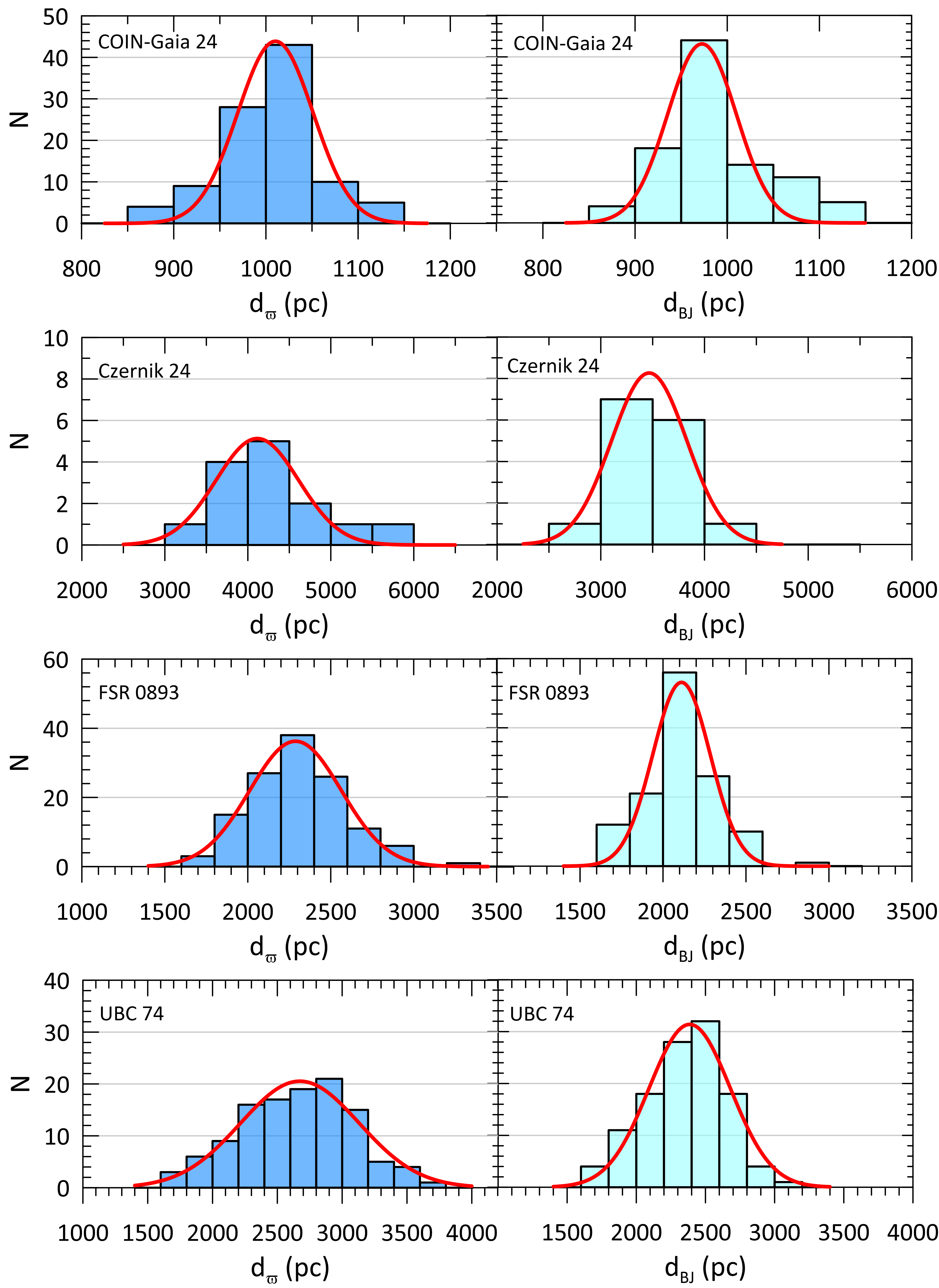}
\caption{Histograms of distances for stars belonging to four OCs ($P\geq 0.5$) with a relative parallax error of $\sigma_{\varpi}/\varpi\leq0.2$, derived from {\it Gaia} DR3 trigonometric parallax measurements (left panels) and from the photogeometric method of \citet{BailerJones2021} (right panels). The red curves represent Gaussian functions fitted to the histograms.}
\label{fig:plx_hist}
\end{figure}

The proper-motion distributions of stars in each OC field, together with their vector directions in equatorial coordinates, are shown in Figure~\ref{fig:VPD_all}. Analysis of the proper-motion diagrams reveals that stars with membership probabilities $P \ge 0.5$ occupy compact regions clearly separated from the general star-field population, with consistent vectorial alignment across the sky. The mean proper-motion components $\langle \mu_{\alpha}\cos\delta, \mu_{\delta} \rangle$ of the most probable members were derived as $(2.503\pm 0.004, -2.958\pm 0.003)$~mas~yr$^{-1}$ for COIN-Gaia~24, $(0.257\pm 0.010, -2.713 \pm 0.007)$~mas~yr$^{-1}$ for Czernik~24, $(-0.333\pm 0.006, -3.183\pm 0.005)$~mas~yr$^{-1}$ for FSR~0893, and $(1.006\pm 0.007, -2.584\pm 0.005)$~mas~yr$^{-1}$ for UBC~74, with the mean values indicated in the corresponding vector point diagrams (VPDs).

The distances of four OCs were calculated using trigonometric parallax data ($\varpi$) from the {\it Gaia} DR3 catalog \citep{Gaia23} and the methods obtained through the Bayesian statistical approach of \citet{BailerJones2021}. In the first method, stars with high cluster membership probabilities ($P\geq 0.50$) were selected by imposing a relative parallax error limit of 0.20, and their distances were determined using the relation $d_{\varpi} ({\rm pc})= 1000/\varpi ({\rm mas})$. In the second method, among the geometric and photogeometric distance estimates produced by \citet{BailerJones2021} for 1.47 billion objects from {\it Gaia} EDR3 data \citep{Gaia21}, only the photogeometric distances ($d_{\rm BJ}$) were adopted in this study, since the photogeometric method based on photometric data allows for more accurate and precise distance determinations in distant OC stars where trigonometric parallax measurements suffer from low S/N ratios. The analyses yielded distances of approximately $1010\pm 41$ pc from trigonometric parallaxes and $973\pm 37$ pc from the \citet{BailerJones2021} estimate for COIN-Gaia~24, based on 99 stars. For Czernik~24, the corresponding values were $4114\pm 500$ pc and $3467\pm 154$ pc from 14 stars, for FSR~0893 they were $2288\pm 277$ pc and $2111 \pm 74$ pc from 127 stars, and for UBC~74 they were $2674 \pm 457$ pc and $2386 \pm 126$ pc from 116 stars (see also Table~\ref{tab:Final_table}). A comparison of the two methods shows that distances derived from trigonometric parallaxes tend to be larger than those obtained by \citet{BailerJones2021}, mainly due to biases inherent in the trigonometric parallaxes, and this discrepancy becomes more pronounced for distant clusters. Therefore, especially for more distant OCs, it is crucial to verify the distance estimates through independent approaches, such as the main-sequence fitting method.  


\section{Determining the structural parameters}

The radial density profile (RDP) analysis of the selected four OCs was carried out using {\it Gaia} astrometric data. Structural parameters of the OCs were obtained from the RDP using cluster member stars, and stellar number density profiles were constructed. Using the method described in Section~\ref{sec:Cluster_Center}, the central equatorial coordinates of the four OCs were calculated in this study (see Table~\ref{tab:Final_table}). During the procedure, stars were counted within concentric annuli of different radii centered on the cluster. For each annulus, the stellar number density was obtained by dividing the number of stars by the area of the annulus, expressed as $\rho(r_i) = N_i / A_i$, where $N_i$ is the number of stars and $A_i$ is the area of the $i^{\rm th}$ annulus. In the analyses, OC distances were adopted from the photogeometric distance ($d_{\rm BJ}$) estimates of \citet{Bailer-Jones21}, as described in Section~\ref{section:cmds} (see Table~\ref{tab:Final_table}).

The structural parameters of the clusters were derived by fitting the King model to the radial density profiles of the analyzed OCs. 
Studying the stellar surface density is particularly useful for investigating the structural properties of an OC, as it describes the variation in the stellar population with projected distance from the cluster center. To represent this distribution, the empirical surface density function proposed by \citet{King62} is commonly applied. The King model describes both the central and outer stellar densities of a cluster through two functional forms: $\rho = \rho_0/(1+(r/r_{\rm c})^2)$ for the inner regions and $\rho = \rho_1 (1/r - 1/r_{\rm t})^2$ for the outskirts. Here, the constants $\rho_0$ and $\rho_1$ act as normalization factors, $r_{\rm c}$ denotes the core radius-defined as the distance from the cluster center where the stellar density decreases to half of $\rho_{\rm 0}$-and $r_{\rm t}$ represents the tidal radius, the point at which the Galactic potential disperses the cluster and the density approaches zero. These relations can be reformulated into a single expression as given in Equation (\ref{equ:king_function}), which encapsulates the King model \citep{Pinfield1998}:
\begin{equation}\label{equ:king_function}
\rho(r) = \rho_0 \left[ \frac{1}{\sqrt{1 + (r/r_c)^2}} - \frac{1}{\sqrt{1 + (r_t/r_c)^2}} \right]^2 + \rho_{\rm bg},
\end{equation}
where $\rho_0$, $r_{\rm c}$, and $r_{\rm t}$ retain the same physical interpretations given above, and $\rho_{\rm bg}$ is the background density in the cluster line of sight. 

The maximum likelihood estimation (MLE) method was implemented to determine the $\rho_0$, $\rho_{\rm bg}$, $r_{\rm c}$, and $r_{\rm t}$ parameters of the model using the log-likelihood function as a measure of goodness of fit. In this approach, the parameter values are selected to maximize the probability of obtaining the observed data given the assumed model. Specifically, the log-likelihood function was defined as:
\begin{equation}\label{equ:chi_square}
\ln \mathcal{L}(\rho_0, \rho_{\rm bg}, r_{\rm c}, r_{\rm t}) = - \sum_i \left(\frac{\rho_{i} - \rho_{i,{\rm Model}}}{\sigma_{\rho_i}}\right)^2,
\end{equation}
where $\rho_{i}$ is the density computed at the $i^{\rm th}$ ring, $\sigma_{\rho_i}$ is the Poissonian uncertainty, and $\rho_{i,{\rm Model}}$ is the density predicted by the King model \citep{King62}. The final estimates of the structural parameters were obtained by maximizing this log-likelihood function, which is equivalent to minimizing the $\chi^2$ statistic. MLE is widely used in astrophysics because it produces estimators that are asymptotically unbiased, consistent, and efficient, while providing a natural framework to quantify uncertainties \citep{Cash1979}. 

\begin{figure*}
\centering
\includegraphics[width=0.99\linewidth]{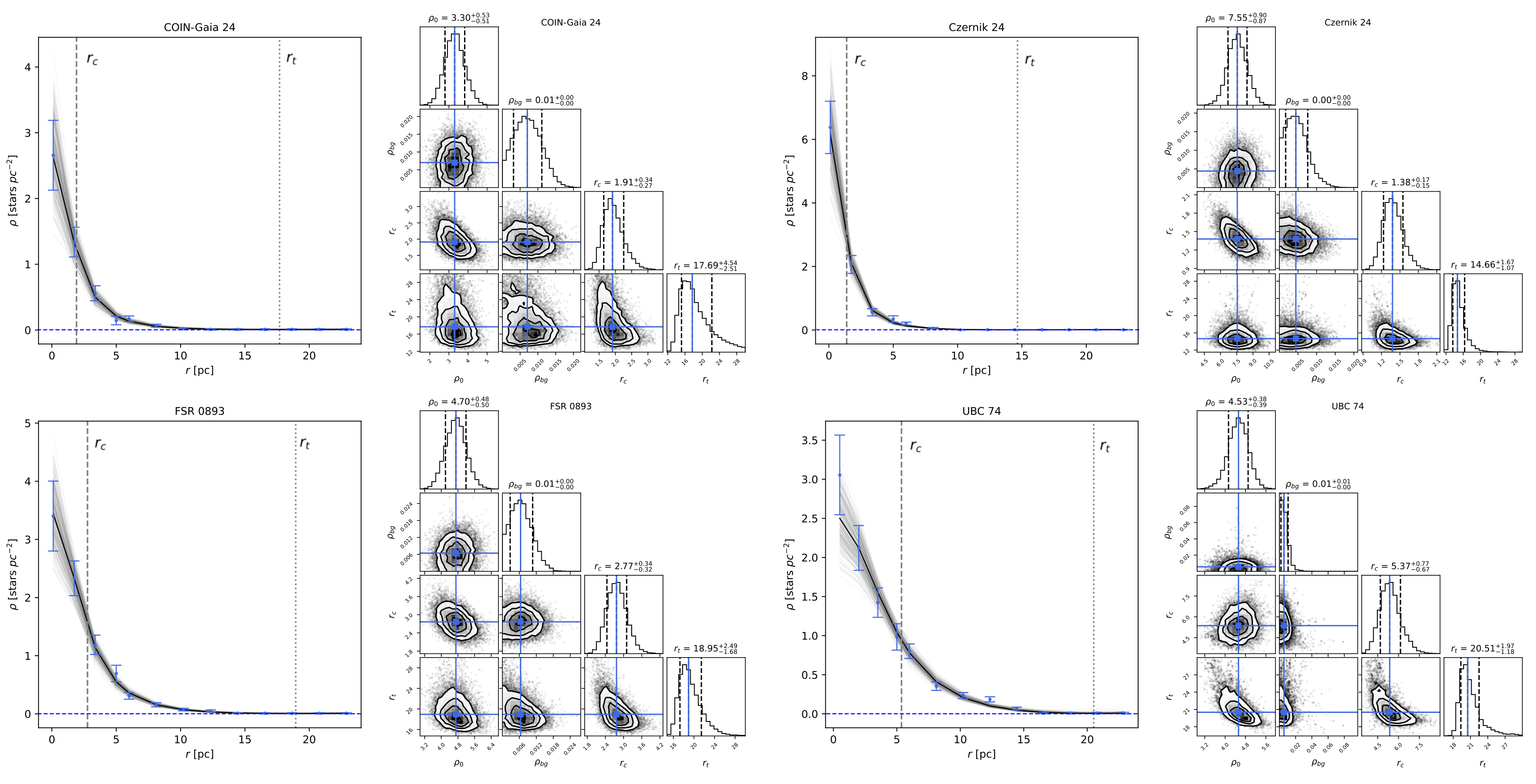}
\caption{RDPs and corner plots of King parameter estimates for the selected four OCs. The structural parameters are expressed as the sample median ($50^{\rm th}$ percentile) together with the ranges defined by the $16^{\rm th}$ and $84^{\rm th}$ percentiles. In the histograms, solid blue vertical lines indicate the median, while the dashed lines mark the $16^{\rm th}$ and $84^{\rm th}$ percentiles. The two vertical lines in the left panels indicate the core radius ($r_{\rm c}$) and the tidal radius ($r_{\rm t}$), while the blue dashed line shows the background density level.}
\label{fig:RDP_Corner}
\end{figure*}

To estimate the best-fitting parameters of the \citet{King62} model, we employed the \texttt{emcee} Python package \citep{Foreman-Mackey2013}, using 32 random walkers and 3000 iterations. Uniform priors were adopted for all parameters. The convergence of the chains was assessed using the \citet{gelman1992} diagnostic. In all cases, we obtained $\hat{R} \leq 1.2$, indicating that the chains had converged. The resulting RDPs for the selected four OCs, along with the best-fit \citet{King62} models, are listed in Table~\ref{tab:radius} and shown in Figure~\ref{fig:RDP_Corner}.The OC \( \rho_{\rm bg} \) values are less than or equal to 0.01 stars pc$^{-2}$ and are therefore omitted from Table~\ref{tab:radius}, but they are only shown in Figure~\ref{fig:RDP_Corner}.

A key structural parameter used to quantify the internal density profile and evolutionary state of an OC is the concentration parameter ($C$). This parameter, as classically defined within the framework of \citet{King62} models, represents the ratio of the cluster's tidal radius ($r_{\rm t}$) to its core radius ($r_{\rm c}$):
\begin{equation}
    C = \left( \frac{r_{\rm t}}{r_{\rm c}} \right)
    \label{eq:concentration}
\end{equation}
The concentration parameter provides a quantitative measure of how strongly the cluster's stellar mass is concentrated towards its gravitational center relative to its tidal boundary, thereby helping to differentiate dynamically evolved, centrally condensed clusters from sparse associations or the surrounding Galactic field population \citep[e.g.,][]{Richstone86}.

\renewcommand{\arraystretch}{1.3}
\begin{table}
\caption{King model parameters determined for four selected OCs. The best-fit King model provides the values of $\rho_0$, $r_{\rm c}$, and $r_{\rm t}$, together with their associated $\chi^2$.}
\label{tab:radius}
\centering
\setlength{\tabcolsep}{4pt}
\begin{tabular}{@{\extracolsep{0.6pt}}lccccc}\hline\hline
\multirow{2}{*}{Cluster} & $\rho_0$ & $r_{\rm c}$ & $r_{\rm t}$ & \multirow{2}{*}{$C$} & \multirow{2}{*}{$\chi^2$} \\ 
\cline{2-4}
 & (stars pc$^{-2}$) & \multicolumn{2}{c}{(pc)} & & \\
\hline
COIN-Gaia~24 & $3.30^{+0.53}_{-0.51}$ & $1.91^{+0.34}_{-0.27}$ & $17.69^{+4.54}_{-2.51}$ & $ 9.26^{+0.62}_{-0.01}$ & 0.047 \\
Czernik~24   & $7.55^{+0.90}_{-0.87}$ & $1.38^{+0.17}_{-0.15}$ & $14.66^{+1.67}_{-1.07}$ & $10.26^{+0.08}_{-0.43}$ & 0.136 \\
FSR~0893     & $4.70^{+0.48}_{-0.50}$ & $2.77^{+0.34}_{-0.32}$ & $18.95^{+2.49}_{-1.68}$ & $ 6.84^{+0.05}_{-0.21}$ & 0.064 \\
UBC~74       & $4.53^{+0.38}_{-0.39}$ & $5.37^{+0.77}_{-0.67}$ & $20.51^{+1.97}_{-1.18}$ & $ 3.82^{+0.16}_{-0.29}$ & 0.237 \\
\hline
\end{tabular}
\end{table}


In this study, the concentration parameters ($C$) for our four target OCs were determined from the best-fit RDPs, yielding values of $9.26^{+0.62}_{-0.01}$ for COIN-Gaia~24, $10.26^{+0.08}_{-0.43}$ for Czernik~24, $6.84^{+0.05}_{-0.21}$ for FSR~0893, and $3.82^{+0.16}_{-0.29}$ for UBC~74. These results indicate that the clusters exhibit a wide range of central concentration levels. COIN-Gaia~24 ($C=9.26$) and Czernik~24 ($C=10.26$) show high central concentrations, characteristic of dynamically evolved systems with well-developed cores formed through long-term internal dynamical processes. FSR~0893 ($C=6.84$) displays an intermediate concentration, suggesting a moderately evolved structure. In contrast, UBC~74 ($C=3.82$) has a relatively low concentration, indicative of a looser system that has likely experienced limited dynamical evolution and core contraction \citep{Elson87, Alzahrani2025a, Alzahrani2025b}. A detailed discussion of these structural properties in relation to the age and dynamical relaxation times of the clusters is provided in Section~\ref{sec:dynamical}.

The structural parameters obtained for the four studied OCs reveal important insights into their stellar distributions. The central stellar densities of these clusters range from 3.30 to 7.55 stars pc$^{-2}$, indicating moderate concentration levels. Core radii span from 1.38 to 5.37 pc, with UBC~74 exhibiting the largest core, suggesting a relatively more extended central region compared to the other clusters. The tidal radii lie between 14.66 and 20.51 pc, reflecting the outer limits of cluster influence under the Galactic tidal field. Overall, these results indicate that the selected clusters possess a range of structural scales, with more compact cores generally associated with higher central densities, while more extended clusters exhibit lower central concentrations. These findings provide a quantitative characterization of cluster structure in the Galactic anti-center region, offering a basis for comparison with clusters in other Galactic environments \citep[e.g.,][]{Bonatto2005, Santos2005}.


\section{Astrophysical Parameters of the Clusters}

This section summarizes two methods applied to obtain fundamental astrophysical parameters for the star clusters COIN-Gaia~24, Czernik~24, FSR~0893, and UBC~74. The first method is based on the simultaneous determination of the fundamental astrophysical parameters of the OCs under investigation using a Bayesian statistical approach. The second method is based on calculating astrophysical parameters by comparing the spectral energy distributions (SED) of the most probable cluster stars with atmospheric models. A detailed description of the basic methodology can be found in previous studies \citep{Bilir06, Bilir10, Bilir16, Yontan15, Yontan19, Yontan21, Yontan22, Ak16, Bostanci15, Bostanci18, Akbulut21, Koc22, ak24, Karagoz25}.

\subsection{Bayesian Inference of Cluster Parameters Using MCMC}
\label{sec: MCMC}
Accurate determination of astrophysical parameters such as age, metallicity, distance modulus, and reddening is essential for understanding the formation and evolution of OCs. To conduct a comprehensive determination of these astrophysical parameters, we employ a Markov Chain Monte-Carlo (\texttt{MCMC}) approach described in \cite{Tanik2025}, in combination with the \texttt{PARSEC} stellar models \citep{Bressan12}.

Overall, the \texttt{MCMC} sampling incorporates a joint posterior distribution over the parameters assuming Gaussian errors. The log-likelihood function, defined following \citet{Tanik2025}, is given by Equation \eqref{equ:log_likelihood}.
\begin{equation}\label{equ:log_likelihood}
\ln \mathcal{L}(\boldsymbol{\theta}) = 
- \sum_{i=1}^{N_{\mathrm{stars}}} \sum_{X}
\left[
\frac{(m_{X,i} - \hat{m}_{X,i})^2}{2\sigma_{X,i}^2}
+ \ln(\sqrt{2\pi}\,\sigma_{X,i})
\right],
\end{equation}
where $\hat{m}_{X,i}$ is the model-predicted magnitude from the stellar grid, $m_{X,i}$ represent the observed magnitude in band $X$ ($G$, $G_{\rm BP}$, $G_{\rm RP}$), $\sigma_{X,i}$ is the uncertainty in band $X_i$, and $\boldsymbol{\theta}$ is the vector of parameters (log age, $A_{\rm G}$, $d$, $Z$). This approach takes into account the observational photometric data for the most probable members in each OC. To explore this posterior, we use the \texttt{emcee} package \citep{Foreman-Mackey2013}, by setting the number of random walkers as the number of the most probable members in each cluster. Then, $5000$ iterations per walker were used to run the sampling. After the algorithm was performed, we obtained the posterior distributions for the cluster parameters, from which the most probable values and 1$\sigma$ uncertainties were extracted. 

\begin{figure*}
\centering
\includegraphics[width=0.85\linewidth]{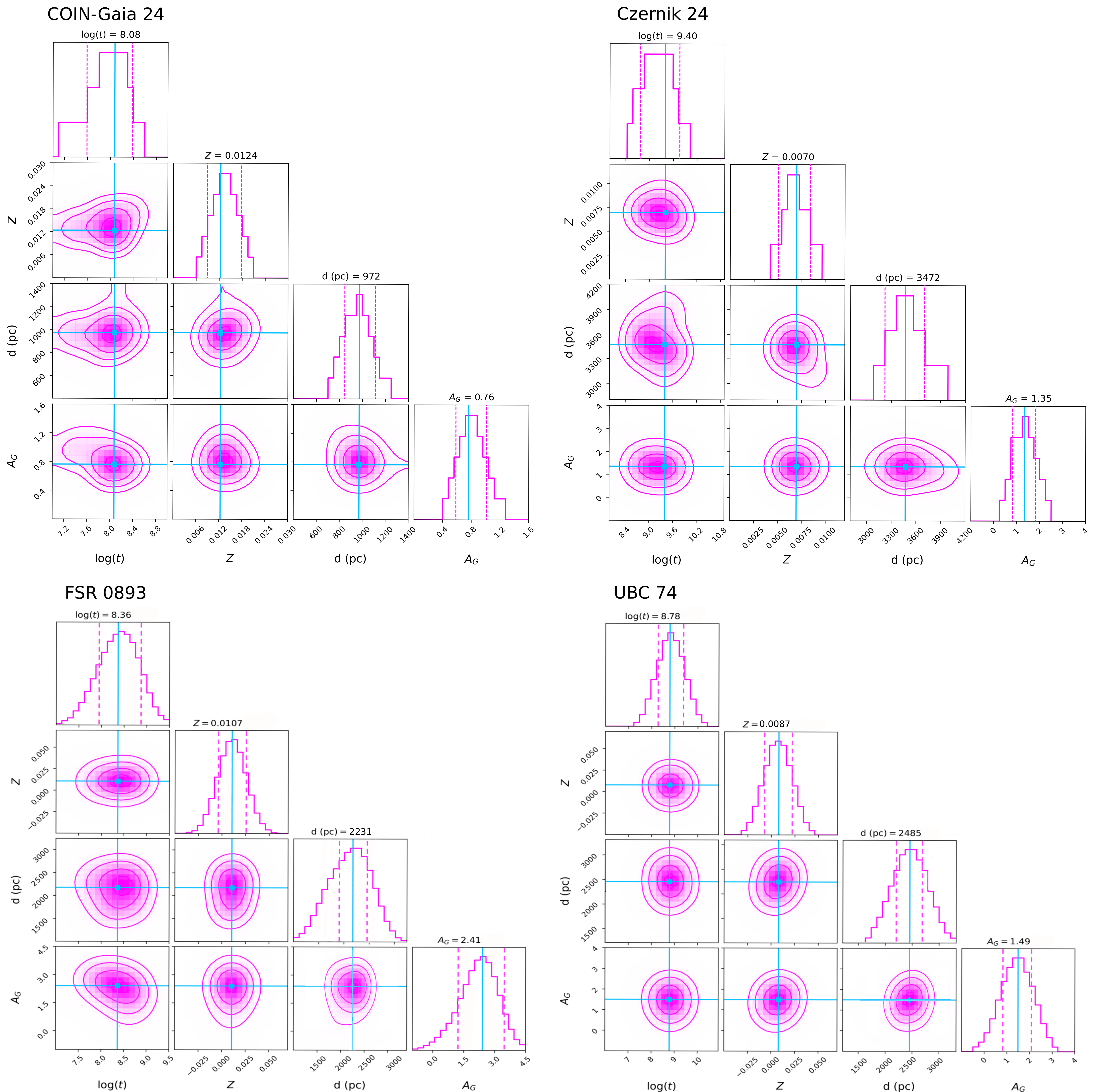}
\caption{Corner plots showing the joint posterior distribution of the astrophysical parameters in each OC after the MCMC sampling. The blue solid lines represent the best parameter, while the dashed pink lines correspond to $16^{\rm th}$ and $84^{\rm th}$ percentiles.}
\label{fig:mcmc_Corner}
\end{figure*}

\begin{figure*}
\centering
\includegraphics[width=0.9\linewidth]{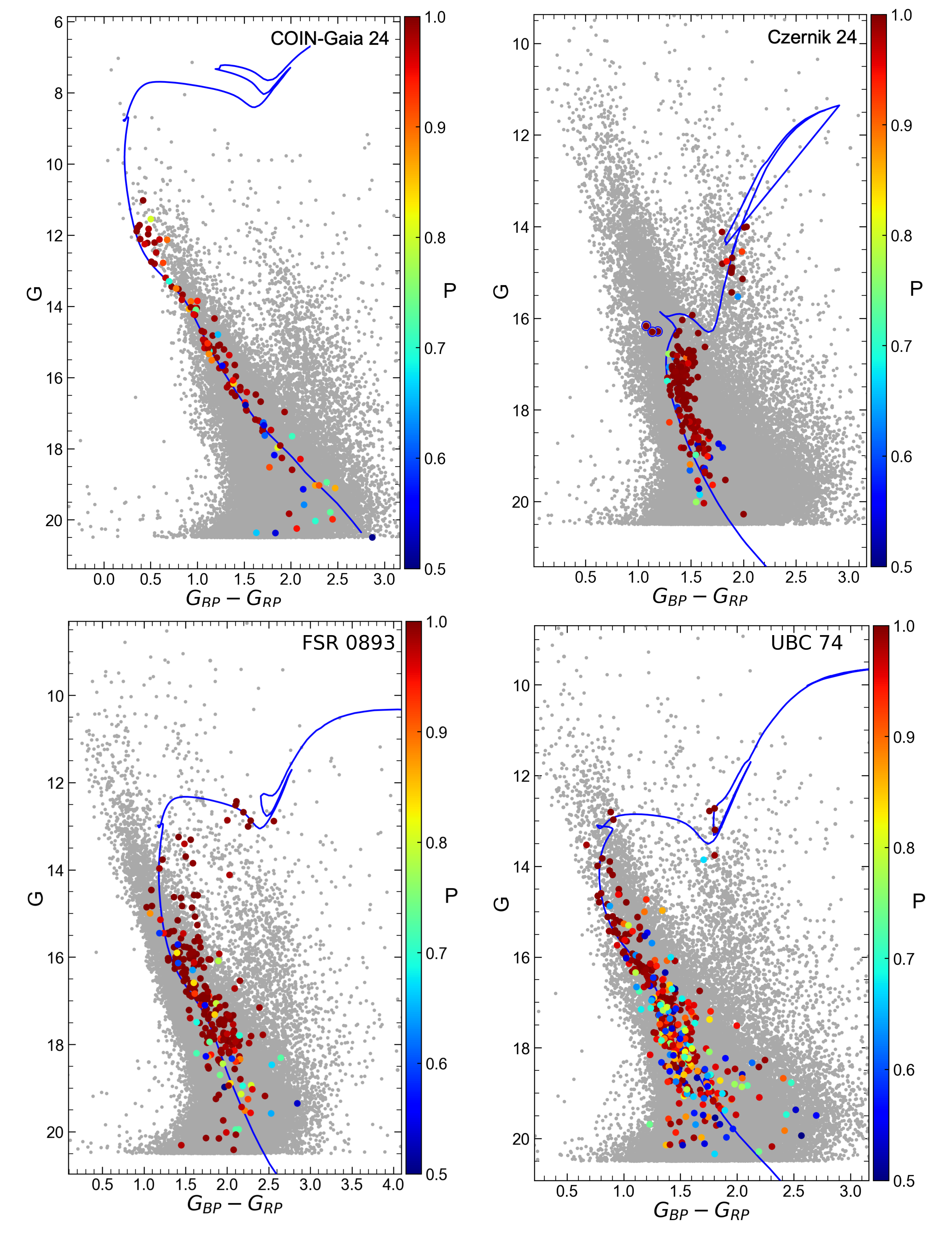}
\caption{CMDs of the selected OCs showing the member stars. Stars with $P \geq 0.5$ are shown in different colours, while field stars are displayed as filled gray circles. The best-fitting \texttt{PARSEC} isochrones (blue curves) and the BSSs are shown in blue circles.}
\label{fig:CMDs_mcmc}
\end{figure*}

Based on this analysis, the parameters were determined as follows in this order (log age, $Z$, distance in pc, and $A_G$ in mag):
8.08, 0.0124, 972, 0.76 for COIN-Gaia~24;
9.40, 0.0070, 3472, 1.35 for Czernik~24;
8.36, 0.0107, 2231, 2.41 for FSR~0893;
and 
8.78, 0.0087, 2485, 1.49 for UBC~74.
Figure~\ref{fig:mcmc_Corner} presents the corner plots of the marginal posterior probability distributions for four OCs, where the lower and upper uncertainties of each parameter correspond to the $16^{\rm th}$ and $84^{\rm th}$ percentiles, respectively. In addition, Figure~\ref{fig:CMDs_mcmc} displays the CMDs of the four OCs, showing the high-probability member stars ($P\geq0.5$) together with the \texttt{PARSEC} isochrones estimated via the MCMC method. Furthermore, the fundamental parameters obtained for the clusters, together with their associated uncertainties, are presented in Table~\ref{tab:Final_table}.

These results provide a comprehensive characterization of the clusters' evolutionary status, allowing for subsequent analyses of their kinematics and structural properties. To account for the effect of metallicity in determining OC ages, the heavy-element content (Z) of each cluster was estimated using the relations provided by Bovy\footnote{\url{https://github.com/jobovy/isodist/blob/master/isodist/Isochrone.py}}, which have been successfully applied in several studies \citep{Gokmen23, Yontan23a, Yontan23b, Yontan23c, Cakmak24}
and are designed for \texttt{PARSEC} stellar isochrones \citep{Bressan12}. In this approach, an intermediate variable $Z_{\rm x}$ is first computed as:
\begin{equation}
Z_{\rm x} = \frac{{\rm Z}}{0.7515 - 2.78 \times {\rm Z}}
\end{equation}
followed by the calculation of the OC iron abundances:
\begin{equation}
{\rm [Fe/H]} = \log \left({\rm Z_{\rm x}} \right) - \log \left( \frac{Z_{\odot}}{1 - 0.248 - 2.78 \times Z_{\odot}} \right)
\end{equation}
In these expressions, $Z_\odot=0.0152$ denotes the solar metallicity. The resulting Z value serves as the adopted heavy-element abundance for the corresponding cluster. For COIN-Gaia~24, Czernik~24, FSR~0893, and UBC~74, the derived metallicities ([Fe/H]) were calculated as $-0.09^{+0.17}_{-0.14}$, $-0.35^{+0.08}_{-0.14}$, $-0.16^{+0.08}_{-0.07}$, and $-0.24^{+0.19}_{-0.18}$ dex, respectively. Also, the parameters derived for the four OCs using the MCMC method are listed in Table~\ref{tab:Final_table}.


\subsection{Spectral Energy Distribution Analysis}

We carried out a comprehensive SED analysis for member stars of the four OCs COIN-Gaia~24, Czernik~24, FSR~0893, and UBC~74, all located in the Galactic anti-center region. The SED modeling was performed using the \texttt{ARIADNE} Python package \citep{Vines2022}, which provides a Bayesian framework for stellar parameter inference from broad-band photometry. The nested sampling algorithm implemented in \texttt{dynesty} \citep{Higson2019, Speagle2020} was employed to efficiently explore the parameter space and compute the Bayesian evidence across competing atmospheric models. The stellar parameters inferred include the effective temperature ($T_{\rm eff}$), surface gravity ($\log g$), metallicity ([Fe/H]), interstellar extinction in the $V$-band ($A_{\rm V}$), stellar mass ($M/M_{\odot}$) and radius ($R/R_{\odot}$), with distances constrained using $Gaia$~EDR3 trigonometric parallaxes from \citet{BailerJones2021}. Extinction values along each line of sight were determined using the dust maps of \citet{Schlafly2011}, ensuring consistent reddening corrections across the stellar samples. Final stellar parameters were obtained through Bayesian model averaging, which combines outputs from ATLAS9/Kurucz stellar atmosphere grids weighted by their statistical evidence, thereby reducing systematic biases \citep{Cinar2025, Elsanhoury2025c, Cinar2026}.

\begin{figure*}
\centering
\includegraphics[width=1\linewidth]{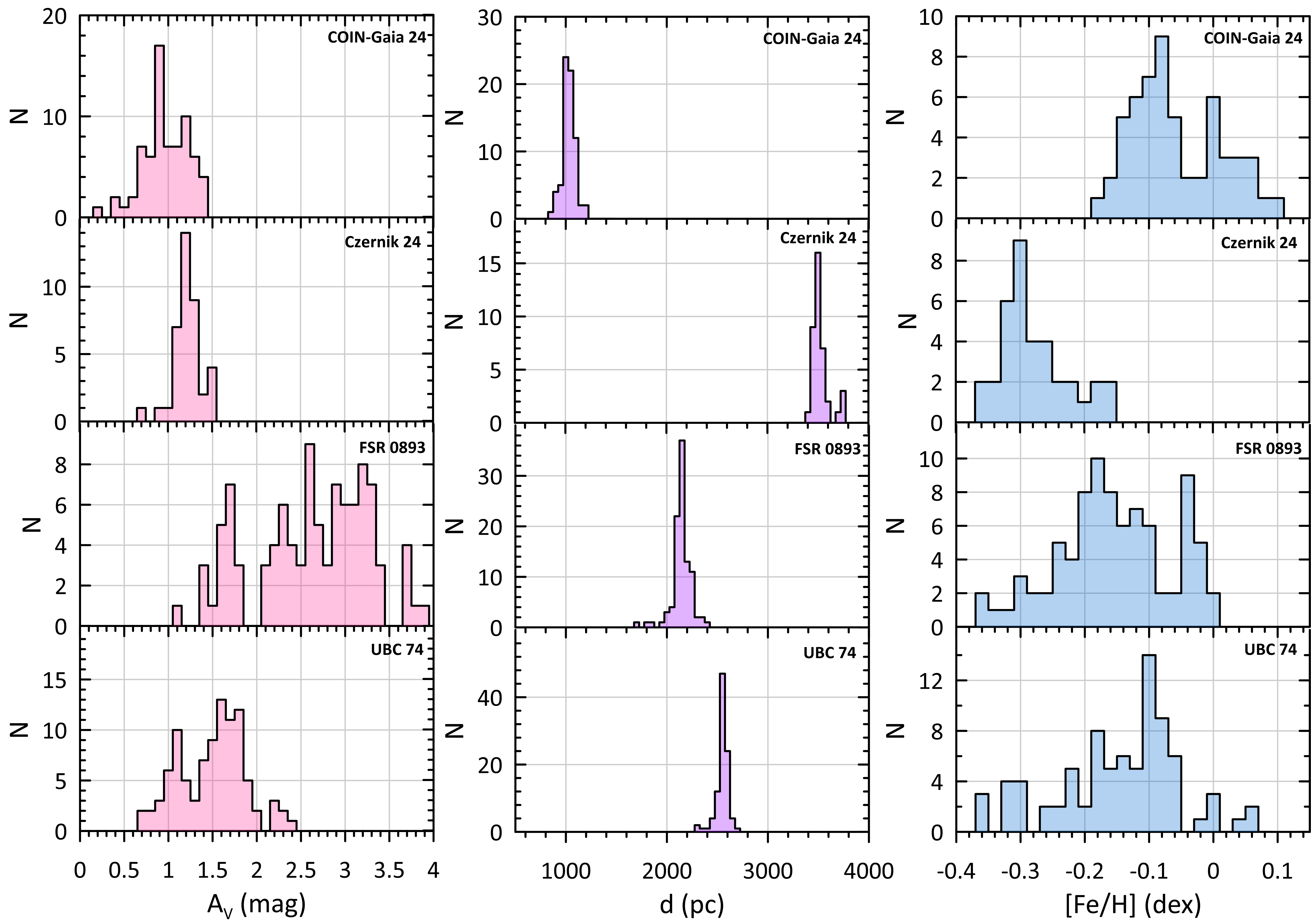}
\caption{Histograms of the $V$-band extinction ($A_{\rm V}$), distance ($d$), and metallicity ([Fe/H]) for 307 stars with SED analysis performed in four OCs are presented.}
\label{fig:SED_results}
\end{figure*}

To construct SEDs, we assembled multi-wavelength photometry from the optical to the infrared, combining $Gaia$~DR3 $G$, $G_{\rm BP}$, and $G_{\rm RP}$ bands \citep{Gaia23} with ground-based surveys such as 2MASS ($J$, $H$, $K_s$; \citealt{Skrutskie2006}) and near- to mid-infrared fluxes from {\it WISE} ($W1$, $W2$, $W3$, $W4$; \citealt{Wright2010}). For brighter stars, additional coverage from Pan-STARRS and APASS was included when available, improving the wavelength baseline and enhancing the reliability of the parameter inference. These heterogeneous data sets were homogenized to absolute flux units following the procedures described in \citet{Cinar2024}.

The total number of cluster members identified in COIN-Gaia~24, Czernik~24, FSR~0893, and UBC~74 was 116, 179, 238, and 387, respectively. To guarantee the robustness of the SED fitting, a strict quality selection was imposed before the analysis. Stars with renormalized unit weight error (\texttt{RUWE}) values exceeding 1.4 in $Gaia$ astrometry \citep{Gaia21, Gaia23} were discarded, as such sources are often affected by unresolved binarity or poor astrometric solutions \citep{Lindegren2021}. Furthermore, a magnitude limit of $G\leq 17$~mag was adopted to mitigate large photometric uncertainties at fainter levels \citep{Cinar2024}. After these constraints, 72 stars were retained in COIN-Gaia~24, 39 in Czernik~24, 100 in FSR~0893, and 96 in UBC~74. In total, 307 stars across the four OCs fulfilled the selection criteria and had sufficiently broad photometric coverage to allow for reliable SED construction.

The histograms of the $V$-band extinction ($A_{\rm V}$), distance ($d$), and metallicity ([Fe/H]) parameters derived from the SED analyses of 307 stars in four OCs are presented in Figure~\ref{fig:SED_results}, with the corresponding mean values summarized in Table~\ref{tab:lumclass}. For COIN-Gaia~24, the mean metallicity is slightly below solar ([Fe/H] = $-0.15\pm 0.02$ dex), with a moderate $V$-band extinction ($A_{\rm V}=0.94\pm 0.03$ mag) and a distance of $1008\pm 8$ pc, indicating a relatively nearby and mildly reddened population. Czernik~24 exhibits a lower metallicity ([Fe/H] = $-0.31\pm 0.01$ dex) and a larger mean distance ($3502\pm 14$ pc), together with higher extinction ($A_{\rm V}=1.56\pm 0.06$ mag), suggesting that it lies in a more distant and obscured region of the Galactic disc. FSR~0893 is found to be slightly more metal-poor ([Fe/H] = $-0.19\pm 0.02$ dex), with a higher $A_{\rm V}$ value of $2.57\pm 0.07$ mag and a distance of $2124\pm 10$ pc, consistent with its location between the solar neighborhood and the outer disc. UBC~74, on the other hand, presents metallicities comparable to FSR~0893 ([Fe/H]=$-0.20\pm 0.01$ dex) but is located somewhat farther away at $2525\pm 6$ pc, with moderate extinction ($A_{\rm V}=1.45\pm 0.04$ mag). The consistency of these parameters across different magnitude limits supports the reliability of the SED-based analysis \citep{Cinar2024}.

\begin{table}
\setlength{\tabcolsep}{5pt}
    \centering
    \caption{Fundamental parameters of the four OCs derived from SED analysis. The table lists the number of member stars ($N$), mean $V$-band extinction ($A_{\rm V}$) mean distance ($d$), and mean metallicity ([Fe/H]). The uncertainties in the parameters represent internal errors, estimated as the standard error ($\sigma/\sqrt{N}$).}
    \begin{tabular}{lcccc}
    \midrule
    \toprule
        Cluster &$N$&  $\langle A_{\rm v}\rangle$ & $\langle d\rangle$ & $\langle {\rm [Fe/H]} \rangle$ \\ 
               && (mag) & (pc) & (dex)    \\ 
        \midrule
        \midrule
COIN-Gaia 24 & 72   &  0.94$\pm$0.03 & 1008$\pm$8 & $-$0.15$\pm$0.02 \\

Czernik 24  & 39  &  1.56$\pm$0.06 & 3502$\pm$14 & $-$0.31$\pm$0.01 \\

FSR 0893   & 100 &  2.57$\pm$0.07 & 2124$\pm$10 & $-$0.19$\pm$0.02 \\

UBC 74      & 96  &  1.45$\pm$0.04 & 2525$\pm$6 & $-$0.20$\pm$0.01 \\
     \bottomrule
        \midrule
    
    \end{tabular} \label{tab:lumclass}
\end{table}

\begin{figure*}
\centering
\includegraphics[width=0.8\linewidth]{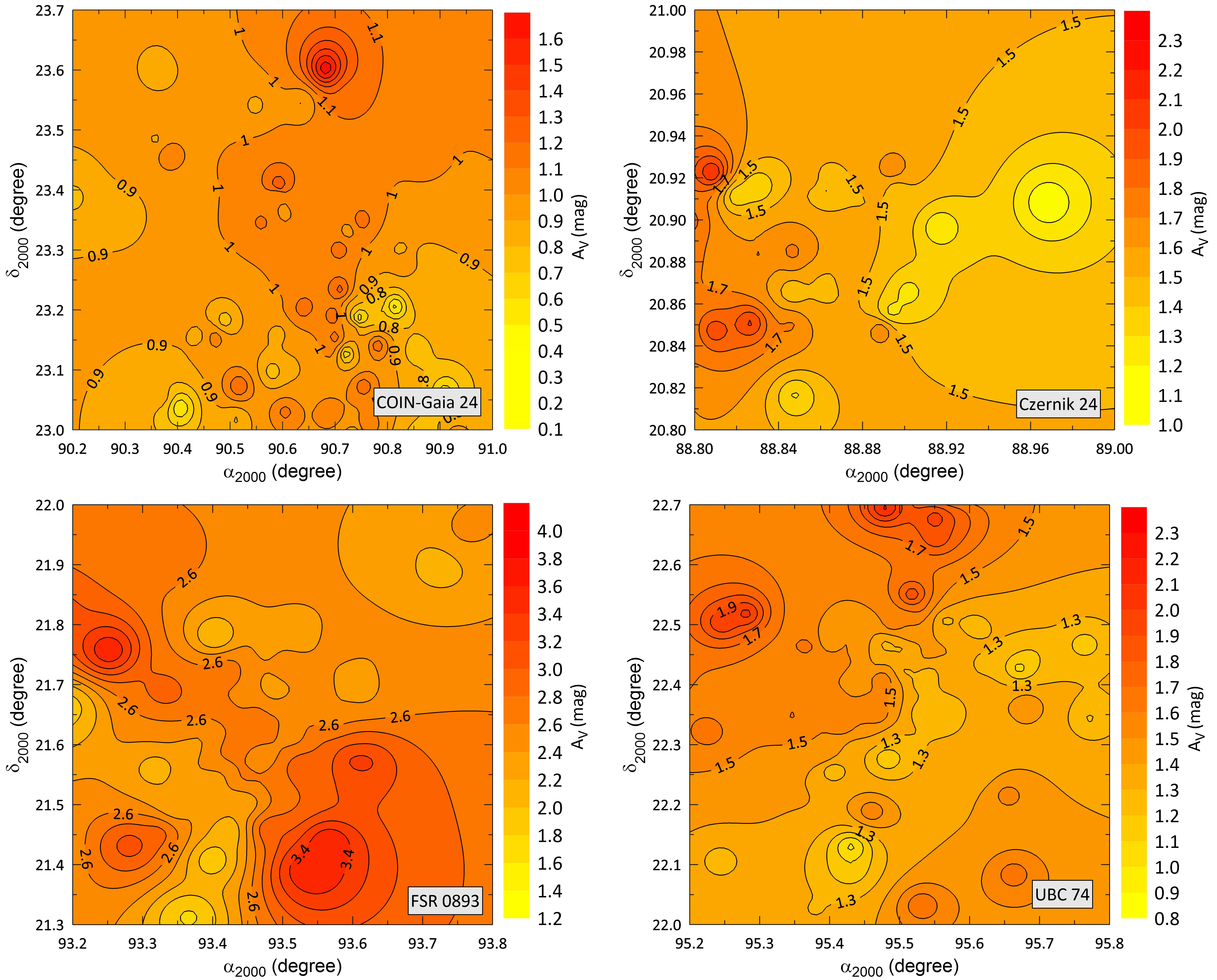}
\caption{Two-dimensional $V$-band interstellar extinction ($A_{\rm V}$) maps for the regions surrounding the four OCs. The contour levels represent variations in $A_{\rm V}$ across the fields, with colour bars indicating the $A_{\rm V}$ values in magnitudes. OC names are marked in the bottom-right corner of each panel.}
\label{fig:SED_kizarma}
\end{figure*}

\begin{figure*}
\centering
\includegraphics[width=0.99\textwidth]{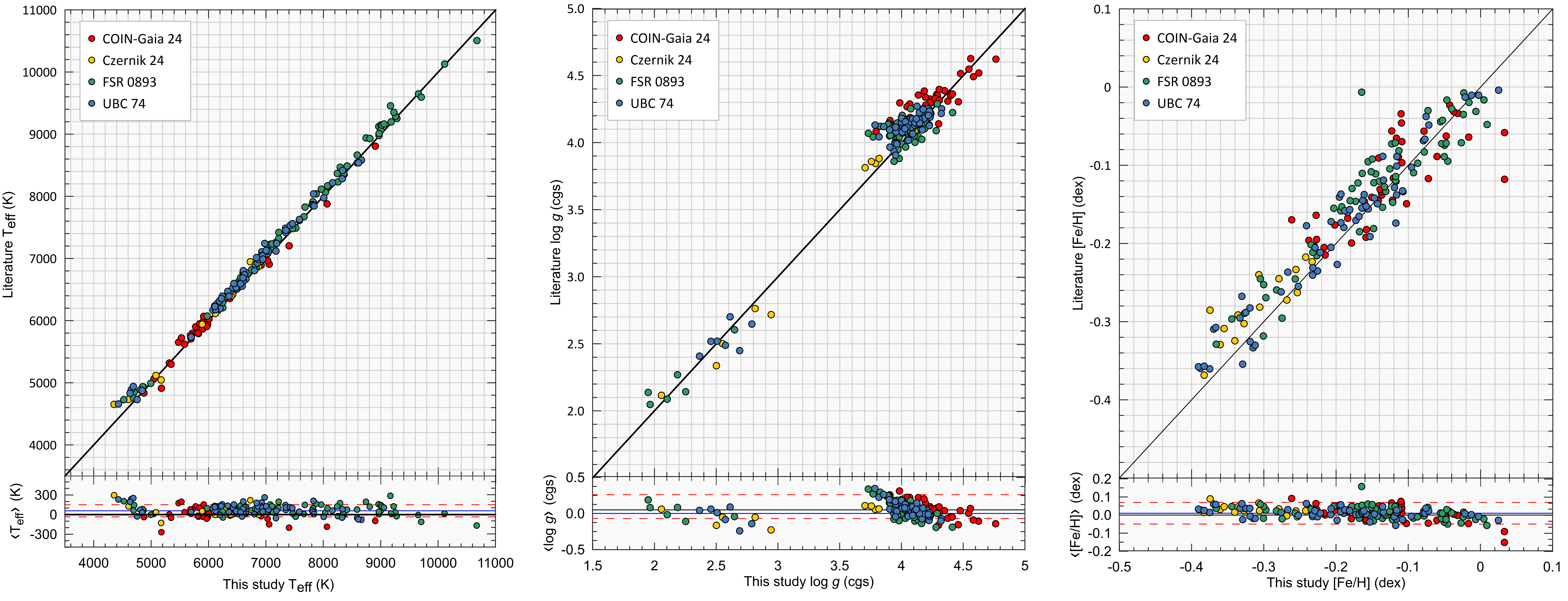}
\caption{Comparison of the atmospheric model parameters derived in this study with those reported in the literature ({\it Gaia} XP) for member stars of the OCs COIN-Gaia~24 (red), Czernik~24 (yellow), FSR~0893 (green), and UBC~74 (blue). Left panel: effective temperature ($T_{\rm eff}$), middle panel: surface gravity ($\log g$), and right panel: metallicity ([Fe/H]). The solid black line represents the one-to-one relation. The lower panels show the residuals between this study and the literature values.}
\label{fig:SED_Lit}
\end{figure*}

Interstellar extinction is a fundamental factor in determining the observed photometric properties of OCs and, therefore, must be carefully characterized in their surrounding regions. Figure~\ref{fig:SED_kizarma} presents the two-dimensional $V$-band extinction maps for the OCs COIN-Gaia~24, Czernik~24, FSR~0893, and UBC~74. In the case of COIN-Gaia~24, extinction values range from approximately $A_{\rm V} \sim 0.1$~mag to 1.5~mag, showing spatial variation across the field. Czernik~24 exhibits higher extinction values (ranging from $A_{\rm V} \sim 1.0$~mag to 1.6~mag and a noticeable gradient toward the eastern side. The extinction map of FSR~0893 reveals the highest obscuration levels among the clusters (approximately $A_{\rm V} \sim 1.2$~mag to 4~mag, with a clear concentration of dust toward the northwestern and southeast regions. UBC~74, on the other hand, displays a wide range of extinction $A_{\rm V} \sim 0.8$~mag to 2.3~mag with localized fluctuations, suggesting patchy interstellar material in its vicinity. Since extinction directly affects both the apparent magnitudes and colour indices of cluster member stars, these spatial variations can introduce systematic uncertainties in isochrone fitting and, consequently, in the derived cluster parameters such as age, distance, and metallicity. These results confirm that interstellar extinction in the vicinity of the OCs is not uniform, indicating the presence of differential reddening across the observed fields.

We cross-matched our sample with the {\it Gaia}~DR3 XP-based stellar parameter catalog produced by \citet{Kordopatis2024}, who transferred spectroscopic labels to 217 million stars using the SHBoost algorithm. Considering the XP quality flags provided for $A_{\rm V}$, $T_{\rm eff}$, $\log g$, [Fe/H], and $M/M_{\odot}$ in their catalog, we retained only the entries with reliable parameter estimates, defined by \texttt{flag = 0} for each quantity. After applying these quality criteria, 237 of the 307 stars analyzed in this study were found to have high-confidence {\it Gaia} DR3 XP counterparts. This subset enabled a robust one-to-one comparison between the atmospheric parameters derived from our SED fitting and those inferred from the {\it Gaia} XP spectra.

When all clusters are examined and listed in Table~\ref{tab:delta_XP}, a general agreement is observed between the SED- and {\it Gaia} DR3 XP-based parameters. Median effective temperature differences across clusters $\langle \Delta T_{\rm eff}\rangle$ are in the range of 16-77~K, respectively, and the median bias for no cluster exceeds 80~K; the total value across clusters is found to be $\langle \Delta T_{\rm eff}\rangle = 58 \pm 93$~K. Median differences for surface gravity $\langle \Delta \log g\rangle$ are at most 0.06~cgs (cluster-median $\langle \Delta \log g\rangle = 0.05 \pm 0.12$~cgs), showing small biases. For metallicity, the values of $\langle \Delta{\rm [Fe/H]}\rangle$ are in the range of 0.00-0.02~dex (cluster-median $\langle \Delta{\rm [Fe/H]}\rangle = 0.01 \pm 0.06$~dex), which is negligible. Generally, the largest biases are observed for $T_{\rm eff}$, followed by $\log g$ and [Fe/H].

\begin{table}
\footnotesize
\centering
\caption{Median differences ($\langle \Delta X \rangle$) with standard deviations ($\sigma_X$) between the SED- and XP-based stellar parameters for the studied clusters.}
\begin{tabular}{lcccc}
\hline \hline
Cluster & $N$ & $\langle \Delta T_{\rm eff} \rangle$& $\langle \Delta \log g \rangle$ & $\langle \Delta {\rm [Fe/H]} \rangle$ \\
& & (K)& (cgs) & (dex)\\
\hline
COIN-Gaia~24 & 61 & $16 \pm 116$ & $0.06 \pm 0.16$ & $0.00 \pm 0.10$ \\
Czernik~24   & 23 & $58 \pm 96$  & $0.04 \pm 0.09$ & $0.02 \pm 0.03$ \\
FSR~0893     & 79 & $77 \pm 75$  & $0.04 \pm 0.12$ & $0.01 \pm 0.04$ \\
UBC~74       & 74 & $64 \pm 81$  & $0.06 \pm 0.10$ & $0.01 \pm 0.03$ \\
\hline
All          &237 & $58 \pm 93$  & $0.05 \pm 0.12$ & $0.01 \pm 0.06$ \\
\hline \hline
\end{tabular}
\label{tab:delta_XP}
\end{table}

\section{Variability, Multiplicity, and Evolutionary Status of Member Stars}
\subsection{Variables and Binaries}

By cross-matching the member stars of COIN-Gaia~24, Czernik~24, FSR~0893, and UBC~74 with the {\it Gaia}~DR3 variability and eclipsing binary catalogs \citep{Mowlavi2023}, we identified some variable stars and eclipsing binary candidates. In COIN-Gaia~24, in particular, two stars are classified as RS CVn-type variables ({\it Gaia} DR3 3424599187150251520 and {\it Gaia} DR3 3424648729596310400), showing photometric modulations typical of active binary systems, while four stars exhibit pulsational variability, with classifications spanning DSCT, GDOR, and SXPHE types ({\it Gaia} DR3 3424606471415073536, {\it Gaia} DR3 3424590769014432768, {\it Gaia} DR3 3424636123869107200, and {\it Gaia} DR3 3424633890485762944). These pulsating stars are located on or slightly above the main sequence in the CMDs and have periods ranging from approximately 0.18 to 0.65 days. The presence of both rotationally modulated and pulsating stars highlights the diversity of stellar variability within these clusters.

In addition to the variability analysis, we also assessed the likelihood of binarity among the cluster members using the {\it Gaia}~DR3 \texttt{RUWE} parameter \citep{Gaia21, Gaia23}. Stars with \texttt{RUWE} $\geq 1.4$ are considered potential binary candidates. Based on this criterion, we identified seven possible binaries in COIN-Gaia~24, one in Czernik~24, four in FSR~0893, and five in UBC~74. These \texttt{RUWE}-selected stars represent additional evidence for the presence of binary systems within the clusters, complementing the eclipsing binary and variability information discussed above.

\subsection{Blue Straggler Stars}

\begin{figure*}
\centering
\includegraphics[width=0.85\linewidth]{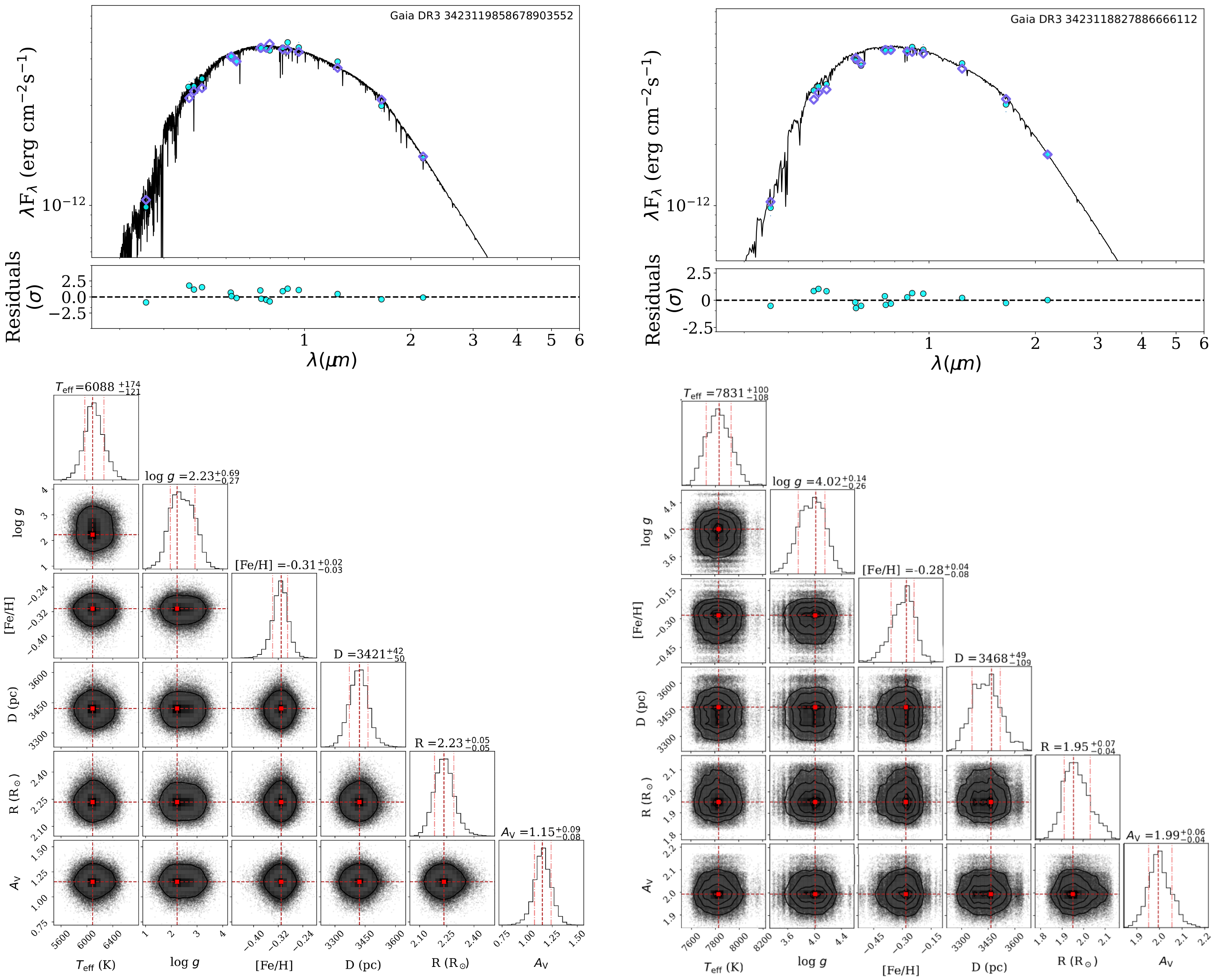}
\caption{Top panels: Best-fitting SED models for two BSS in Czernik 24, {\it Gaia} DR3 3423119858678903552 (left) and {\it Gaia} DR3 3423118827886666112 (right). Black curves show the best-fitting models, while cyan pluses and circles indicate the observed photometric measurements. Blue diamonds mark the synthetic photometry derived from the models. Bottom panels: Posterior distributions from the corresponding SED fittings. Red points denote the mean values of the posterior samples, and red vertical lines represent the 16$^{\rm th}$, 50$^{\rm th}$, and 84$^{\rm th}$ percentiles, respectively.}
\label{fig:SED_BSS}
\end{figure*}

We identified a total of three blue straggler stars (BSS) among the member stars of four OCs. In particular, Czernik~24 hosts two confirmed BSSs ({\it Gaia} DR3 3423118793528219008 and {\it Gaia} DR3 3423119858678903552) and one uncertain BSS (BS?, {\it Gaia} DR3 3423118827886666112), while other OCs do not show any BSSs in our sample \citep{Jadhav21}. These stars are located above the main-sequence turn-off in the CMD of the cluster, appearing bluer and hotter than the bulk of the main-sequence population. Their spatial distribution shows a mild central concentration within the cluster radius, consistent with dynamical formation channels such as stellar collisions and binary interactions, which are expected to be more frequent in dense cluster environments.

The confirmed BBS in Czernik 24 ({\it Gaia} DR3 3423119858678903552) exhibits SED-derived parameters consistent with classical BSS characteristics. The effective temperature ($T_{\rm eff}=6088^{+174}_{-121}$ K) places the star in the lower F-type regime, while the surface gravity ($\log g=2.23^{+0.69}_{-0.27}$ cgs) is no longer on the main sequence but has evolved beyond it \citep{Ferraro2006}. The stellar radius ($R=2.23^{+0.05}_{-0.05} ~R_{\odot}$) and luminosity ($L=6.18^{+1.24}_{-0.98}~L_{\odot}$) further support this interpretation, suggesting a post–post-mass-transfer or mildly evolved stage of a BSS. The metallicity ([Fe/H]$=-0.31^{+0.02}_{-0.03}$ dex) aligns with the cluster’s mean chemical composition, while the distance ($d=3421^{+42}_{-50}$ pc) and the interstellar extinction ($A_{\rm V}=1.15^{+0.09}_{-0.08}$ mag) are consistent with Czernik 24. Overall, these parameters confirm the star’s nature as a BSS and its likely membership in the Czernik 24.

The blue straggler candidate in Czernik 24 ({\it Gaia} DR3 3423118827886666112), though uncertain, exhibits SED-derived parameters, as shown in  Figure~\ref{fig:SED_BSS}, that are partially consistent with a classical BSS profile. The effective temperature ($T_{\rm eff}=7831^{+100}_{-108}$ K) places the star at the lower end of the F-type regime, while the surface gravity ($\log g=4.02^{+0.14}_{-0.26}$ cgs) falls within the range expected for main-sequence or BSSs. The stellar radius ($R=1.95^{+0.07}_{-0.04}~R_{\odot}$) and luminosity ($L=12.92^{+1.93}_{-1.51}~L_{\odot}$) are significantly larger than those of typical main-sequence stars, suggesting that the object may lie in the BSS or subgiant/early giant domain. The metallicity ([Fe/H]$=-0.28^{+0.04}_{-0.08}$ dex) is consistent with the cluster’s mean chemical composition, and both the distance ($d=3468^{+49}_{-109}$ pc) and $V$-band extinction ($A_{\rm V}=1.99^{+0.06}_{-0.04}$ mag) are in agreement with the parameters of Czernik~24.

SED analyses of two BSS candidates in Czernik 24 yield consistent fundamental parameters. These results demonstrate that SED fitting can complement spectroscopic studies, although it could not be performed for {\it Gaia} DR3 3423118793528219008 due to insufficient photometric magnitudes. These results indicate that the source shares several characteristics with BSSs, although its evolutionary status and cluster membership require confirmation through high-resolution spectroscopy and complementary radial velocity and astrometric analyses.


\section{Kinematic and Orbital analysis}

\begin{table*}
\centering
\setlength{\tabcolsep}{2.pt}
\renewcommand{\arraystretch}{0.65}
\caption{Radial velocity data for stars with high membership probabilities in the four OCs, including their {\it Gaia} DR3 source names, equatorial coordinates, $G$-band apparent magnitudes, radial velocities, \texttt{RUWE} values, and membership probabilities.}
{\scriptsize
    \begin{tabular}{cccccclcc}\\
          \hline
          \hline
Order & Gaia Source ID          & $\alpha_{\rm J2000}$ & $\delta_{\rm J2000}$ &   $G$  & RV            & \texttt{RUWE} & $P$ \\
      &                         & ($^\circ$)           &	($^\circ$)        &  (mag) & (km s$^{-1}$) &      &   \\
  \hline
  \hline
    \multicolumn{8}{c}{COIN-Gaia 24} \\
\hline
\hline
01 &  3424598392578265344 & 90.69850 & 23.18969 & 11.550$\pm$0.003 & -16.71$\pm$14.17 & 0.947 & 0.804 \\
02 &  3424595510658249344 & 90.74450 & 23.18671 & 11.026$\pm$0.003 &   4.98$\pm$8.44  & 0.996 & 0.995 \\
03 &  3424593582214897408 & 90.72051 & 23.12760 & 12.211$\pm$0.003 &  -4.00$\pm$17.85 & 1.142 & 0.990 \\
04 &  3424600389741073664 & 90.70916 & 23.30179 & 14.824$\pm$0.003 &   0.55$\pm$10.54 & 1.003 & 0.993 \\
05 &  3424594376786886656 & 90.77999 & 23.14109 & 11.977$\pm$0.003 & -30.01$\pm$13.85 & 1.124 & 0.993 \\
06 &  3424595785536132608 & 90.81507 & 23.20606 & 11.720$\pm$0.003 &  19.27$\pm$13.24 & 1.078 & 0.994 \\
07 &  3424592452641562240 & 90.75388 & 23.07329 & 12.222$\pm$0.003 &  -3.89$\pm$21.29 & 1.059 & 0.992 \\
08 &  3424590494136501376 & 90.58223 & 23.09797 & 14.073$\pm$0.003 & -29.63$\pm$6.39  & 1.158 & 0.861 \\
09 &  3424623101528100352 & 90.82424 & 23.29805 & 13.862$\pm$0.003 & -18.07$\pm$8.77  & 1.007 & 0.898 \\
10 &  3424619352019720704 & 90.86977 & 23.24494 & 14.030$\pm$0.003 &  14.52$\pm$9.97  & 1.063 & 0.990 \\
11 &  3424648248558625792 & 90.56013 & 23.34598 & 14.102$\pm$0.003 & -14.13$\pm$11.44 & 1.059 & 0.739 \\
12 &  3424651444014292736 & 90.49761 & 23.37056 & 11.789$\pm$0.003 & -10.79$\pm$16.24 & 1.072 & 0.992 \\
13 &  3424626812379770624 & 90.86948 & 23.45653 & 12.133$\pm$0.003 &   0.47$\pm$6.88  & 1.133 & 0.899 \\
14 &  3424577506155687808 & 90.92971 & 22.98599 & 14.723$\pm$0.003 &   2.46$\pm$7.33  & 1.027 & 0.981 \\
15 &  3424589291545709824 & 90.40413 & 23.03735 & 13.818$\pm$0.003 &   7.71$\pm$8.03  & 1.115 & 0.971 \\
16 &  3424678283267858304 & 90.72860 & 23.55577 & 11.827$\pm$0.003 &  -7.10$\pm$12.15 & 1.135 & 0.979 \\
17 &  3424688694268307072 & 90.54881 & 23.54105 & 14.695$\pm$0.003 &   7.95$\pm$17.35 & 1.067 & 0.987 \\
18 &  3424657289467486464 & 90.29972 & 23.35891 & 14.228$\pm$0.003 &   0.61$\pm$7.42  & 1.047 & 0.960 \\
19 &  3424670517966722048 & 90.36851 & 23.60234 & 13.500$\pm$0.003 & -14.52$\pm$12.87 & 0.910 & 0.892 \\
  \hline
  \hline
    \multicolumn{8}{c}{Czernik 24}   \\
  \hline
  \hline

01 & 3423864502927505664 & 88.74765 & 20.89072 & 14.809$\pm$0.003 & 20.12$\pm$6.11   & 1.058 & 0.995 \\
02 & 3423113850021054848 & 88.78628 & 20.88326 & 14.545$\pm$0.003 & -29.49$\pm$3.47  & 0.996 & 0.925 \\
03 & 3423119759895492864 & 88.79780 & 20.89944 & 15.143$\pm$0.003 & 22.52$\pm$8.52   & 1.071 & 0.996 \\
04 & 3423118522944888064 & 88.89322 & 20.85531 & 14.892$\pm$0.003 & 16.48$\pm$4.97   & 0.980 & 0.996 \\
05 & 3423119278859107072 & 88.87849 & 20.90921 & 14.017$\pm$0.003 & 22.44$\pm$2.27   & 1.026 & 0.997 \\
06 & 3423120000413620224 & 88.85878 & 20.91612 & 14.681$\pm$0.003 & 22.14$\pm$3.22   & 0.984 & 0.997 \\
07 & 3423118763463062272 & 88.86291 & 20.87117 & 15.007$\pm$0.003 & 23.24$\pm$6.10   & 1.037 & 0.996  \\
08 & 3423118729103333248 & 88.83517 & 20.88108 & 14.757$\pm$0.003 & 20.52$\pm$5.80   & 1.105 & 0.946 \\
09 & 3423118763463061376 & 88.85941 & 20.87617 & 14.982$\pm$0.003 & 27.64$\pm$5.37   & 1.035 & 0.994 \\
10 & 3423118729103333120 & 88.84101 & 20.87604 & 14.003$\pm$0.003 & 19.30$\pm$2.60   & 1.097 & 0.997  \\
11 & 3423118827886641664 & 88.84609 & 20.87766 & 14.957$\pm$0.003 & 19.75$\pm$4.57   & 1.041 & 0.996 \\
   \hline
   \hline
    \multicolumn{8}{c}{FSR 0893}  \\
  \hline
\hline
01 & 3376873682357151744 & 93.41418 & 21.63478 & 12.872$\pm$0.003 & 39.97$\pm$0.68   & 1.013 & 0.999 \\
02 & 3376873682357153408 & 93.40436 & 21.63597 & 13.006$\pm$0.003 & 38.72$\pm$0.72   & 1.124 & 0.998 \\
03 & 3376873888515582848 & 93.39322 & 21.64944 & 12.679$\pm$0.003 & 38.78$\pm$0.99   & 1.078 & 0.998 \\
04 & 3376875327325751424 & 93.44448 & 21.67762 & 12.430$\pm$0.003 & 39.84$\pm$0.79   & 0.958 & 0.998 \\
05 & 3376868837634009344 & 93.56624 & 21.66074 & 12.519$\pm$0.003 & 37.74$\pm$1.04   & 1.098 & 0.998 \\
06 & 3376859590567532928 & 93.60659 & 21.59990 & 12.882$\pm$0.003 & 39.36$\pm$0.83   & 1.113 & 0.994 \\
07 & 3376882787687784704 & 93.47853 & 21.75158 & 12.863$\pm$0.003 & 40.33$\pm$1.22   & 1.182 & 0.987 \\
   \hline
   \hline
        \multicolumn{8}{c}{UBC 74}    \\
  \hline
\hline
01 & 3377161479525506688 & 95.45119 & 22.40985 & 13.197$\pm$0.003 & 41.51$\pm$2.23   & 1.067 & 0.998 \\
02 & 3377161376446861056 & 95.45217 & 22.39658 & 12.781$\pm$0.003 & 41.64$\pm$0.64   & 0.965 & 0.998 \\
03 & 3377161440868410368 & 95.43345 & 22.41162 & 12.724$\pm$0.003 & 43.04$\pm$1.62   & 1.286 & 0.997 \\
04 & 3377162991353629696 & 95.42680 & 22.47481 & 14.976$\pm$0.003 & 14.27$\pm$9.51   & 1.043 & 0.863 \\
05 & 3377167904796550144 & 95.50748 & 22.47298 & 14.706$\pm$0.003 & 63.36$\pm$4.87   & 1.015 & 0.989 \\
06 & 3377209892396508928 & 95.40295 & 22.47121 & 13.760$\pm$0.003 & 30.69$\pm$1.84   & 1.026 & 0.989 \\
07 & 3377168248393903232 & 95.60365 & 22.49659 & 13.856$\pm$0.003 & 26.13$\pm$2.21   & 1.194 & 0.670 \\
08 & 3376831660396619904 & 95.22762 & 22.32070 & 14.658$\pm$0.003 & 20.14$\pm$4.01   & 1.133 & 0.959 \\
09 & 3376766376894157696 & 95.42803 & 22.09255 & 14.732$\pm$0.003 & 31.49$\pm$9.92   & 0.935 & 0.935 \\
      \hline
      \hline
    \end{tabular}%
  \label{tab:RV_info}%
  }
\end{table*}%

Radial velocity data are crucial for conducting detailed kinematic and orbital dynamical analyses of OCs. In this study, the mean radial velocities of the four OCs were estimated using {\it Gaia} DR3 measurements \citep{Gaia23}. To ensure a reliable membership selection, only stars with a cluster membership probability $P\geq 0.5$ were included in the analysis. COIN-Gaia~24, Czernik~24, FSR~0893, and UBC~74, the numbers of stars with available radial velocity data among those with high cluster membership probabilities were determined to be 19, 11, 7, and 9, respectively. The {\it Gaia} source IDs, equatorial coordinates, $G$ apparent magnitude \texttt{RUWE} values, and cluster membership probabilities of these stars are presented in Table~\ref{tab:RV_info}. According to the table, the stars with radial velocity measurements have cluster membership probabilities of $P\geq0.67$, and their $G$ apparent magnitudes lie within the range of 11-15. An examination of the \texttt{RUWE} values shows that none of the stars exceed 1.4, indicating that no prominent binary stars are present in the compiled list. The mean radial velocity ($\langle V_{\rm R}\rangle$) for each OC was computed using a weighted mean approach, following the methodology outlined by \citet{Soubiran18}. The resulting values yielded $\langle V_{\rm R}\rangle=-4.99\pm 3.28$ km~s$^{-1}$ for COIN-Gaia 24, $\langle V_{\rm R}\rangle=15.57 \pm 5.29$ km~s$^{-1}$ for Czernik 24, $\langle V_{\rm R}\rangle=39.30\pm 0.31$ km~s$^{-1}$ for FSR 0893 and $\langle V_{\rm R}\rangle= 39.80\pm 2.75$ km~s$^{-1}$ for UBC 74.

To derive the Galactic orbits of the OCs, we adopted a realistic three-component Milky Way potential, consisting of a Miyamoto-Nagai disk \citep{Miyamoto75}, a Hernquist bulge \citep{Hernquist1990}, and a logarithmic halo \citep{Binney2008}, as implemented in the \texttt{MWPotential2014} (MW14) model within the \texttt{galpy} package \citep{Bovy15}. This potential provides an analytic and axisymmetric representation of the Galactic disk, bulge, and halo components, widely used in studies of stellar and cluster orbits. The space-velocity components and orbital parameters of the four OCs were calculated using their equatorial coordinates ($\alpha$, $\delta$), mean radial velocities ($\langle V_{\rm R}\rangle$), mean proper-motion components ($\langle\mu_{\alpha}\cos\delta, \mu_{\rm \delta}\rangle$), and distances ($d_{\rm iso}$) along with corresponding uncertainties derived in this study. We adopted a Galactocentric distance of $R_{\rm gc}=8$ kpc, a circular rotation speed of $V_{\rm rot}=220$ km~s$^{-1}$ \citep{Bovy15, Bovy12}, and a vertical distance of the Sun from the Galactic plane of $27\pm 4$ pc \citep{Chen_2000}.

To trace the past orbits of the OCs, their orbits were extended until a closed orbital path was obtained, allowing for an accurate determination of the orbital properties of each OC, integrated backward in time over 3 Gyr with a temporal resolution of 1.5 Myr. This analysis yielded the orbital parameters including the apogalactic ($R_{\rm a}$) and perigalactic ($R_{\rm p}$) distances, orbital eccentricity ($e$), maximum vertical distance from the Galactic plane ($Z_{\rm max}$), space velocity components ($U$, $V$, $W$), and orbital periods ($T_{\rm P}$). The space-velocity components $(U, V, W)$ of the four OCs COIN-Gaia24, Czernik~24, FSR~0893, and UBC~74 were determined as (7.09, -16.79, 3.32), (-9.28, -40.57, -18.23), (-34.73, -34.09, -17.78), and (-33.86, -38.96, -0.76) km~s$^{-1}$, respectively. It should be noted that these velocities are heliocentric and therefore include the contribution of the solar motion. To eliminate this effect, corrections to the Local Standard of Rest (LSR) were applied using the solar motion values reported by \citet{Coskunoglu11}. The corrected velocity components $(U, V, W)_{\rm LSR}$, along with the total space velocity $S_{\rm LSR}$ and Galactic orbital parameters, are listed in Table~\ref{tab:Final_table}. 

In order to study the three-dimensional spatial distribution of the four OCs, their projected Cartesian coordinates in the Galactic reference frame were derived from the heliocentric distances of their member stars. These coordinates, expressed as $(X, Y, Z)_{\odot}$, were obtained using the relations $X=d_{\rm iso}\cos l \cos b$, $Y=d_{\rm iso} \sin l \cos b$, and $Z=d_{\rm iso} \sin b$, where $d_{\rm iso}$ is the heliocentric distance and $(l, b)$ denote the Galactic longitude and latitude of the OC center \citep[see also,][]{Canbay2025, Caliskan2025}. In this framework, $X$ corresponds to the axis directed toward the Galactic center, $Y$ indicates the axis aligned with the Sun’s rotational direction, and $Z$ represents the axis pointing out of the Galactic plane toward the North Galactic Pole. The Galactocentric distance ($R_{\rm gc}$) of each OC was then calculated with the expression $R_{\rm gc} = \sqrt{R_{\rm 0}^{2} + (d_{\rm iso} \cos b)^2 - 2 R_{\rm 0} d_{\rm iso} \cos l \cos b}$ \citep{Cinar2025}, where $R_{\rm 0} = 8$ kpc is adopted as the solar distance from the Galactic center \citep{Majewski93}. The distances of the four OCs in the Cartesian coordinates together with their Galactocentric distances are presented in Table~\ref{tab:Final_table}.

This study employs the concept of the traceback early orbital radius ($R_{\rm teo}$), introduced by \citet{Akbaba24}, which characterizes the past orbital state of a stellar system by integrating its orbit within a static Galactic potential. In contrast to the birth radius ($R_{\rm birth}$), which seeks to pinpoint the exact formation site of a cluster, $R_{\rm teo}$ provides information about its earlier orbital location in the Galaxy. The estimation of $R_{\rm teo}$ was performed together with other kinematic and dynamical parameters.

\begin{figure*}
\centering
\includegraphics[width=0.88\textwidth]{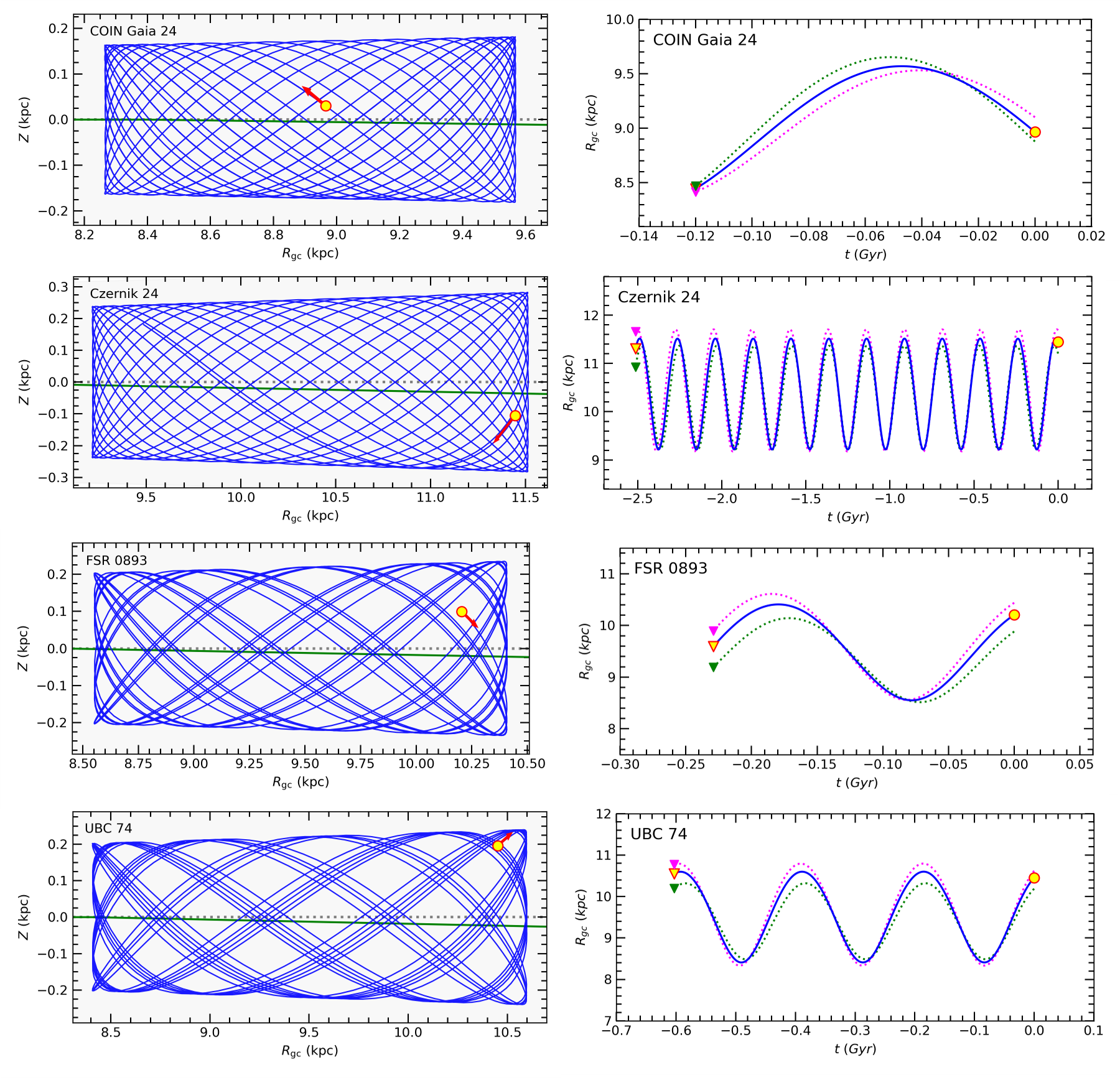}
\caption{Galactic orbits of the four OCs are shown in the $Z \times R_{\rm gc}$ plane (left) and the $R_{\rm gc} \times t$ plane (right). Filled yellow triangles and circles denote the clusters’ birthplaces and present-day positions, respectively. The upper and lower sets of input parameters for the orbits and birthplaces are indicated by pink and green filled triangles, along with the corresponding dashed lines. Red arrows represent the motion vectors of the OCs. Dotted and solid green lines in the panels show the Galactic plane and warp plane, respectively.}
\label{fig:galactic_orbits}
\end{figure*}

Figure~\ref{fig:galactic_orbits} shows the Galactic orbits of the four OCs. The left and right panels display their motions in the $R_{\rm gc}\times Z$ plane \citep[e.g.][]{Tasdemir23, Tasdemir2025, TasdemirCinar2025, Tasdemir2026, Elsanhoury2025}, while the right-hand plots further show the temporal evolution of their Galactocentric distances in the $R_{\rm gc}\times t$ plane \citep[e.g.][]{Yontan23c, Yucel24, Yucel25, Haroon2025}. In  Figure~\ref{fig:galactic_orbits}, present-day and early orbital positions of the OCs are indicated by yellow-filled circles and triangles, respectively. Dashed pink and green lines, along with the corresponding triangles, represent the orbital paths and early orbital locations derived by accounting for uncertainties in the input parameters. The upper panels of Figure~\ref{fig:galactic_orbits} show that four OCs originated outside the solar circle and have remained confined to this region throughout their orbital evolution. Orbital dynamics analyses reveal that the birthplaces of COIN-Gaia~24, Czernik~24, FSR~0893, and UBC~74 were located at Galactocentric distances of 8866, 10238, 9388, and 9379 pc, respectively (see also, Table~\ref{tab:Final_table}). These calculations show that the birthplaces of the studied OCs were located at distances different from their present-day Galactic radii. According to the calculations, the differences between the current distances of COIN-Gaia~24, Czernik~24, FSR~0893, and UBC~74 from the Galactic center and their birthplaces are found to be 0.1, 1.21, 0.82, and 1.07 kpc, respectively. The fact that these distance differences are not very large indicates that the present-day positions of the studied OCs can be reliably used in the analysis of the metallicity gradient.

The fact that the four OCs investigated in this study lie in opposite directions with respect to the Galactic center implies that they are subject to both the Galactic warp and flare. The sinusoidal expression models the vertical displacement of an object due to the warp:
\begin{equation}
Z_{\rm warp}(R,\phi)=\gamma_{\rm w}\,[R - R_{\rm warp}] \sin(\phi - \phi_{\rm w}),
\end{equation}
where the warp onset radius is taken as $R_{\rm warp} = 8.4$ kpc and the line-of-nodes angle as $\phi_{\rm w} = 5^{\circ}$ \citep{Derriere2001}. These parameter values are consistent with Besançon-like Galaxy models \citep{Robin12} and observational studies based on 2MASS and \textit{Gaia}, which typically place the warp onset near $8-9$ kpc and find a small ($\sim$ few degrees) phase offset relative to the Sun--Galactic center line \citep{Lopez2002, Reyle2009, Chrobakova2022, Uppal2024}. The adopted warp amplitude parameter, $\gamma_{\rm w}=0.09$ kpc$^{-1}$, was selected so that the resulting vertical displacements in the range $R_{\rm gc} \approx 9-12$ kpc match the values reported in the literature. This choice is meant as a simplified but observationally compatible parametrization, rather than a newly optimized fit to the global disc structure. Using this warp prescription, the vertical positions of the OCs were corrected through:
\begin{equation}
Z_{\rm cor} = Z - Z_{\rm warp}(R,\phi).
\end{equation}

To investigate the flare of the Galactic disc, a radius-dependent scale height $h_{\rm z}(R)$ is adopted. For $R_0 = 8\,\mathrm{kpc}$ a local thin-disc scale height of $h_{\rm z,0} = 0.30$ kpc \citep[e.g.][]{Bilir2008, Iyisan2025} is assumed, while for $R > R_0$ the relation:
\begin{equation}
h_z(R) = h_{z,0}\,\left[1 + \gamma_{\mathrm{f}}(R - R_0)\right]
\end{equation}
is applied, with $\gamma_{\mathrm{f}} = 0.05\,\mathrm{kpc^{-1}}$. This mild outward increase in scale height aligns well with results from 2MASS, SDSS, and \textit{Gaia}-based studies of the outer-disc thickening \citep{Lopez2002, Momany2006, Lopez2014, Chrobakova2022, Uppal2024}.

Orbits integrated within the \texttt{MW14} potential yield the maximum vertical amplitude, the orbital eccentricity, and the mean Galactocentric radius. Using the warp-corrected coordinates, an effective vertical amplitude $Z_{\rm max, wf}$ is derived. The fraction of the orbital period for which the condition $|Z - Z_{\rm warp}|<0.1$ kpc holds is defined as the ``warp matching fraction'' ($\rm W_{mf}$). Additionally, the median value of $|Z|/h_z$ along each orbit is computed to quantify the clusters' typical height relative to the flared thin disc. For their present-day positions, the ratio $|Z_{\rm now}|/h_{\rm z}$ and the local flare factor ($f_{\rm fac}$):
\begin{equation}
f_{\rm fac} = \frac{h_{\rm z}(R_{\rm now}) - h_{\rm z,0}}{h_{\rm z,0}}
\end{equation}
are evaluated. These warp and flare parameters are listed in Table~\ref{tab:tab:warp-flare} and shown in the $R_{\rm gc}\times Z$ panels in Figure~\ref{fig:galactic_orbits}.


The derived Galactic orbital parameters of the four OCs were employed to determine their Galactic population membership. Following the classification scheme of \citet{Schuster12}, stars with $V_{\rm LSR}$ velocities greater than $-50$ km s$^{-1}$ are considered part of the thin disc, those with $-180 < V_{\rm LSR}~({\rm km~s^{-1}})\leq -50$ are attributed to the thick disc, and velocities below $-180$ km s$^{-1}$ indicate halo membership. Based on the calculated $V_{\rm LSR}$ values, four OCs are identified as members of the Galactic thin-disc population. Their Galactic orbits exhibit low eccentricities, not exceeding 0.15, and vertical distance from the Galactic plane limited to about $Z_{\rm max}=300$ pc. These kinematic and Galactic orbital characteristics further corroborate their classification as thin-disc clusters \citep{Plevne15, Guctekin19}.

\begin{table}
  \setlength{\tabcolsep}{1pt}
  \renewcommand{\arraystretch}{1}
  \small
  \centering
  \caption{Warp and flare parameters for the four OCs.}
    \begin{tabular}{lccccccc}
    \hline
                     & MW14  & $\rm W_{\rm cor}$ &   &   &       &       &  \\
    \hline
    Cluster & $Z_{\rm max}$  & $Z_{\rm max}$  & $\Delta Z_{\rm max}$ & $W_{\rm mf}$ & $|Z|/h_{\rm z}$ & $|Z_{\rm now}|/h_{\rm z}$ & $f_{\rm fac}$ \\
                 & (pc)  & (pc) & (\%)  & (\%)  &       &       & (\%) \\
            \hline
    COIN-Gaia 24 & 181 & 274 & 52 & 43  & 0.39 & 0.10 & ~~5\\
    Czernik 24   & 283 & 550 & 95 & 26  & 0.56 & 0.30 & 17 \\
    FSR 0893     & 235 & 408 & 74 & 33  & 0.48 & 0.30 & 11 \\
    UBC 74       & 240 & 435 & 81 & 32  & 0.49 & 0.58 & 12 \\
      \hline
            \end{tabular}%
  \label{tab:tab:warp-flare}%
\end{table}%


\section{Investigating the Dynamical Behavior of the Clusters}
\subsection{Luminosity and Mass Functions}

In this study, we derived the luminosity functions (LFs) for four selected OCs, specifically COIN-Gaia~24, Czernik~24, FSR~0893, and UBC~74. Our analysis is predicated on the catalogs constructed from the {\it Gaia} DR3 astrometric and photometric data. To estimate from $G$-band apparent magnitudes to $G$-band absolute magnitudes ($M_{\rm G}$), we employed the standard distance modulus equation, correcting for interstellar extinction. Utilizing the distance ($d$) and colour excess $E(G_{\rm BP}-G_{\rm RP})$ values derived from the isochrone fitting procedure (Section~4.1), we computed $M_{\rm G}$ for each member star via the relation: $M_{\rm G} = G - 5 \log d +5-A_{\rm G}$. The total extinction in the $G$-band, $A_{\rm G}$, was calculated using $A_{\rm G} = 1.8626 \times E(G_{\rm BP}-G_{\rm RP})$, based on the extinction coefficients derived by \citet{Cardelli89} and \citet{ODonnell94} for the {\it Gaia} passbands \citep{Canbay2023}.

\begin{figure}
\centering
\includegraphics[width=0.5\linewidth]{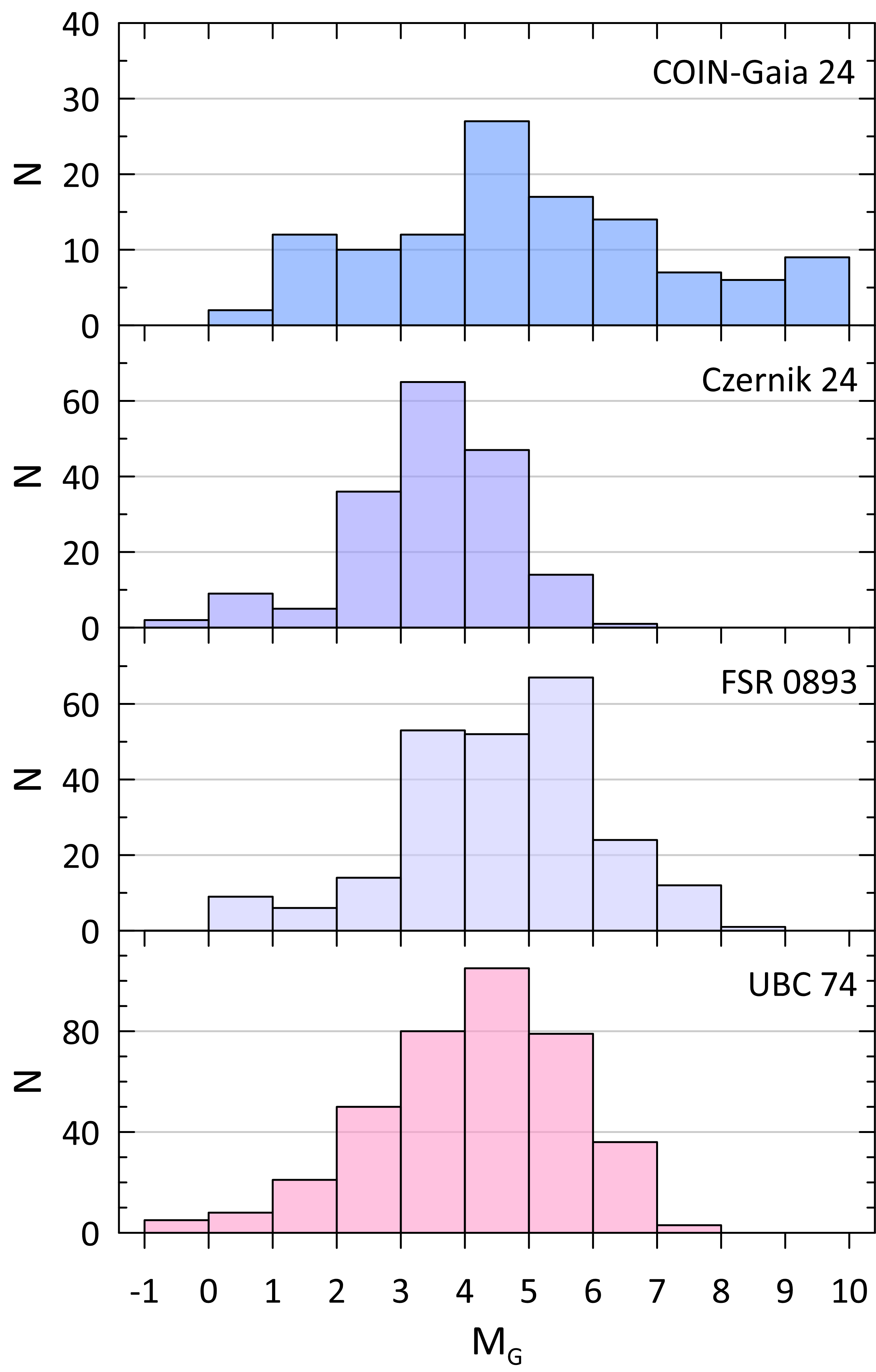}
\caption{The histograms of LFs for four OCs.}
\label{fig:luminosity_functions}
\end{figure}

This transformation yielded absolute magnitude ranges of $0 <M_{\rm G}~({\rm mag})\leq 10$ for COIN-Gaia~24, $-1 <M_{\rm G}~({\rm mag})\leq 7$ for Czernik~24, $0<M_{\rm G}~({\rm mag})\leq 9$ for FSR~0893, and $-1<M_{\rm G}~({\rm mag})\leq 8$ for UBC~74. The resulting LFs, constructed by binning the star counts per unit $M_{\rm G}$ interval, are presented as histograms in Figure~\ref{fig:luminosity_functions}. Inspection of panels in Figure~\ref{fig:luminosity_functions} reveals distinct features for each cluster. The star counts peak at $M_{\rm G} \approx 5$~mag for COIN-Gaia~24, $M_{\rm G} \approx 4$~mag for Czernik~24, $M_{\rm G} \approx 6$~mag for FSR~0893, and $M_{\rm G} \approx 5$~mag for UBC~74. After these peaks, the LFs exhibit a decline towards fainter magnitudes.

As a complementary analysis, we derived the present-day mass functions (PDMFs) for the four OCs. Stellar masses are not direct observables; thus, they were inferred by leveraging the theoretical \texttt{PARSEC} stellar evolutionary models \citep{Bressan12}, which were employed to determine cluster ages and distance moduli. The \texttt{PARSEC} model grids provide a direct mapping between stellar mass and absolute magnitudes across various photometric systems for a given age and metallicity. For each OC, we utilized the best-fitting isochrone to establish a precise mass-luminosity relation (MLR). A sixth-order polynomial function was fitted to the theoretical $G$-band absolute magnitude ($M_{\rm G}$) versus mass ($M/M_{\odot}$) data from the corresponding \texttt{PARSEC} isochrone. These MLRs were then applied to convert the calculated $M_{\rm G}$ absolute magnitudes of the selected MS stars into stellar mass estimates.

To define the selected MS stars, we considered stars from the turn-off point down to fainter magnitudes within each cluster. Specifically, for COIN-Gaia~24, stars with apparent $G$ magnitudes between 11 and 19 (covering in absolute magnitudes $0<M_{\rm G}~(\rm mag) \leq 8$) were included, totalling 102 stars. For Czernik~24, the selected range was $G=16.5-19.5$ mag ($2.5<M_{\rm G}~(\rm mag)\leq 5$), comprising 140 stars. FSR~0893 stars spanned $G=14.5-18.5$ mag ($2.2<M_{\rm G}~(\rm mag)\leq 6.1$) with 191 stars, and UBC~74 included stars in $G=15-18.5$ mag ($1.7<M_{\rm G}~(\rm mag)\leq 5.2$), totalling 262 stars.

The resulting mass intervals spanned by the analyzed MS stars are: $0.55<M/M_{\odot}\leq3.1$ for COIN-Gaia~24, $0.9< M/M_{\odot}\leq1.35$ for Czernik~24, $0.7< M/M_{\odot}\leq1.7$ for FSR~0893, and $0.8< M/M_{\odot}\leq1.9$ for UBC~74. These mass ranges correspond directly to the apparent $G$-band magnitude limits noted earlier. We emphasize that the {\it Gaia} photometry is expected to be complete - even at the faint end of these magnitude ranges - as demonstrated using the Besançon Galaxy model in Section~\ref{sec:Completeness}. This completeness mitigates significant biases in the PDMF analysis, particularly concerning the potential undercounting of low-mass stars, thereby ensuring the robustness of our derived PDMF slopes.

\begin{figure}
\centering
\includegraphics[width=0.5\linewidth]{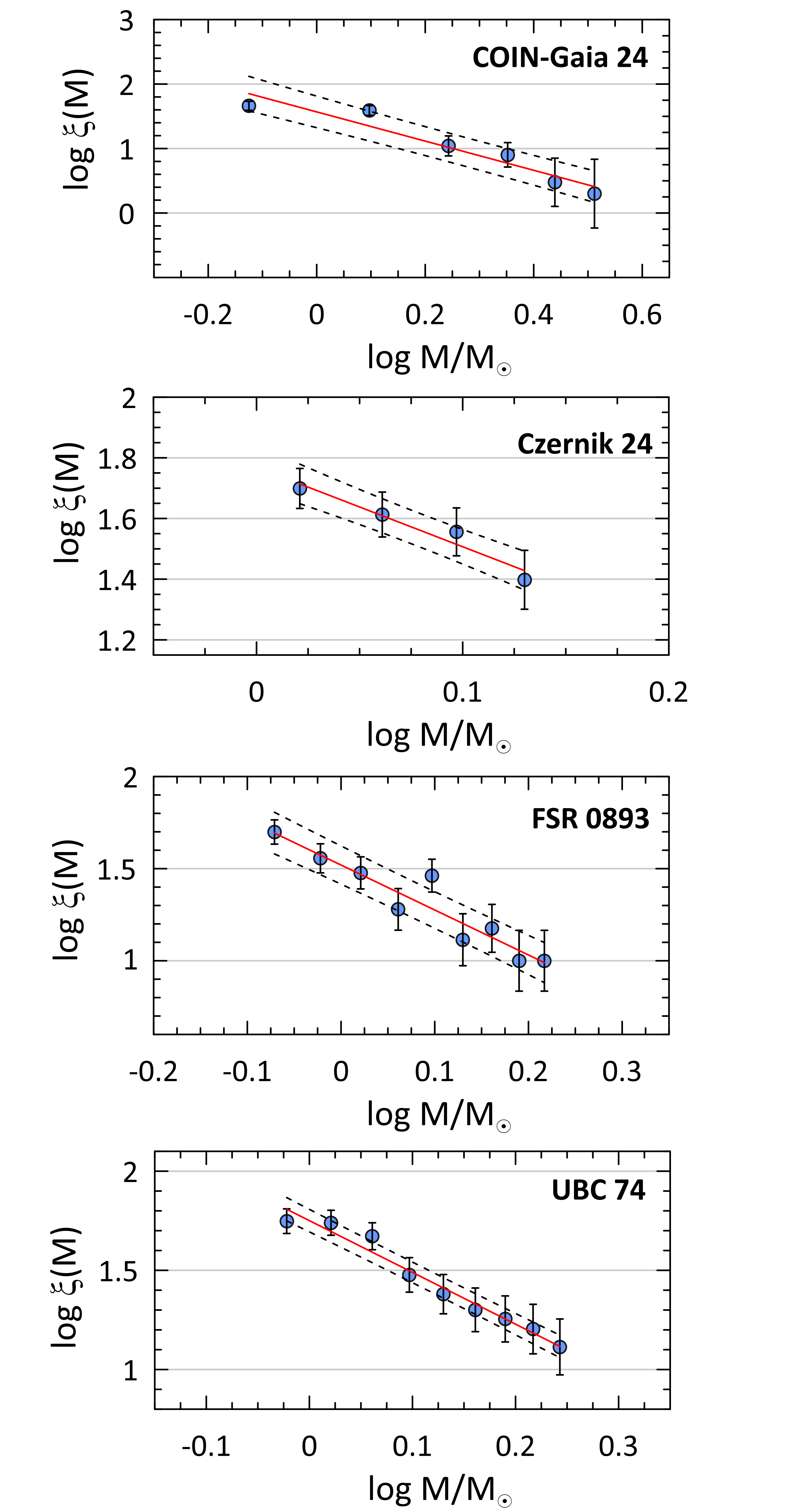}
\caption{PDMFs for four OCs. The mass distributions (blue circles) are shown along with the red lines applied to the distribution. The black lines indicate the prediction levels ($\pm 1\sigma$) of the linear fit.}
\label{fig:mass_functions}
\end{figure}

To compute the PDMFs, the stellar mass distribution for each cluster was generated. The logarithmic count of stars in each mass bin was then calculated. These distributions are shown in Figure~\ref{fig:mass_functions}. The uncertainties associated with the star counts in each mass bin were calculated assuming Poisson statistics ($1/\sqrt{N}$). For the determination of the PDMF slopes ($\Gamma$), we utilized the logarithmic form of the classical power-law relation introduced by \citet{Salpeter55}:
\begin{equation}
    \log \left( \frac{dN}{dM} \right) = -(1 + \Gamma) \times \log M + C
    \label{eq:salpeter}
\end{equation}
where $dN$ is the number of stars within a mass interval $dM$, $M$ is the central mass of that interval, $C$ is a constant, and $\Gamma$ is the slope of the PDMF. A linear regression fit based on Equation~\ref{eq:salpeter} was applied to the mass distributions, represented by the blue solid lines in Figure~\ref{fig:mass_functions}. The PDMF slopes derived from this fitting procedure are $\Gamma=1.16\pm 0.35$ for COIN-Gaia~24, $\Gamma=1.62\pm 0.47$ for Czernik~24, $\Gamma=1.44\pm 0.32$ for FSR~0893, and $\Gamma=1.61\pm 0.18$ for UBC~74. These results are discussed in Section~\ref{sec:discuss}, where they are compared against the canonical \citet{Salpeter55} value of $\Gamma = 1.35$.

\subsection{The Dynamical State of Mass Segregation}
\label{sec:dynamical}
Mass segregation, the phenomenon wherein the most massive stars in a stellar system preferentially concentrate towards the cluster core, stands as a fundamental signature of dynamical evolution. This process is an inevitable consequence of two-body relaxation, driven by cumulative gravitational interactions among cluster members \citep{Raboud98}. These interactions gradually steer the cluster towards a state of kinetic energy equipartition, a process during which high-mass stars systematically lose kinetic energy to their lower-mass counterparts. This energy exchange causes the more massive stars to sink towards the gravitational potential well (the core), while the lower-mass stars gain velocity and migrate towards the cluster's periphery or halo \citep[e.g.,][]{Hillenbrand98, Bisht20}.

The characteristic timescale for this dynamical ``sorting'' is the half-mass relaxation time ($T_{\rm E}$). This metric quantifies the time required for a star's velocity to be significantly altered by stellar encounters. $T_{\rm E}$ is critically dependent on the number of cluster members ($N$), the half-mass radius ($R_{\rm h}$, in pc), and the mean stellar mass ($\langle m \rangle$). Following \citet{Spitzer71}, it is classically expressed as:
\begin{equation}
T_{\rm E} = \frac{8.9 \times 10^{5} N^{1/2} R_{\rm h}^{3/2}}{\langle m\rangle^{1/2}\log(0.4N)}
\label{eq:t_relax}
\end{equation}

A cluster's observed dynamical state is determined by the ratio of its chronological age ($\tau_{\text{age}}$) to its relaxation time. If $\tau_{\text{age}} / T_{\rm E} \gg 1$, the cluster is considered dynamically ``relaxed'' or ``evolved'', and any observed mass segregation is likely the product of this long-term dynamical process. Conversely, if $\tau_{\text{age}} / T_E \lesssim 1$, the cluster is dynamically young, and any observed central concentration of massive stars is more likely ``primordial''-a relic of the initial conditions of star formation rather than subsequent evolution \citep{Allison09, Haroon2025}. This picture is further complicated by external factors such as the Galactic tidal field and internal processes like stellar feedback, which can accelerate or disrupt the cluster's evolution \citep{Kruijssen12}.

We have computed the dynamical relaxation times for the four OCs in our sample. For COIN-Gaia~24, we identified $N = 116$ stars with a total mass of $M_{\text{cl}} = 131 M_{\odot}$, yielding a mean mass $\langle m \rangle = 1.129 M/M_{\odot}$. For Czernik~24, we found $N = 179$ stars, $M_{\text{cl}} = 203 M_{\odot}$, and $\langle m \rangle = 1.134 M/M_{\odot}$. For FSR~0893, $N = 238$ stars, $M_{\text{cl}} = 268 M_{\odot}$, and $\langle m \rangle = 1.126 M/M_{\odot}$. Lastly, for UBC~74, we identified $N = 387$ members, $M_{\text{cl}} = 459 M_{\odot}$, and $\langle m \rangle = 1.186 M/M_{\odot}$. The half-mass radii ($R_{\rm h}$) were determined to be $3.08^{+0.66}_{-0.44}$~pc for COIN-Gaia~24, $2.38^{+0.28}_{-0.22}$~pc for Czernik~24, $3.86^{+0.40}_{-0.40}$~pc for FSR~0893, and $5.63^{+0.68}_{-0.52}$~pc for UBC~74.

Substituting these values into Equation~(\ref{eq:t_relax}), we estimate the relaxation times as $T_{\rm E} = 29.3^{+10.0}_{-6.0}$~Myr, $22.1^{+3.1}_{-2.4}$~Myr, $49.6^{+9.7}_{-7.4}$~Myr, and $98.2^{+18.2}_{-13.3}$~Myr for COIN-Gaia~24, Czernik~24, FSR~0893, and UBC~74, respectively (see also Tablo~\ref{tab:Final_table}). Comparing these timescales with the cluster ages derived in this study (see Table~\ref{tab:Final_table}), we find the ratios $\tau_{\text{age}}/T_{\rm E}$ to be approximately $4.1$, $113.7$, $4.6$, and $6.1$, respectively. Notably, the substantial ratio obtained for Czernik 24 ($\tau_{\text{age}} / T_{\rm E}=113.7$) may also reflect its location at a Galactocentric distance of about 11.5 kpc, where external tidal forces and disc-shocking effects are significantly weaker. This environment likely allows the cluster to retain its members longer and reach a highly relaxed dynamical state with minimal external perturbations.

\begin{table*}
  \centering
  \setlength{\tabcolsep}{10pt}
  \renewcommand{\arraystretch}{1}
  \caption{Basic parameters obtained for four OCs.}
  \medskip
  {\small}
        \begin{tabular}{lrrrr}
\hline
Parameter & COIN-Gaia 24 & Czernik 24 & FSR 0893 & UBC 74 \\
\hline
$\alpha_{\rm J2000}$ (Sexagesimal)& 06:02:46.32 & 05:55:23.52 & 06:13:48.96 & 06:21:50.64 \\
$\delta_{\rm J2000}$ (Sexagesimal) &$+$23:12:10.8 & $+$20:52:33.6 & $+$21:36:28.8 & $+$22:25:08.4 \\
$l_{\rm J2000}$ (Decimal)  & 186.89279 & 188.05847&189.52650 &189.69013 \\
$b_{\rm J2000}$ (Decimal)  &$+$0.41579   &$-2$.22632&$+$1.87372  &$+$3.89684  \\
Cluster members ($P\geq0.5$)              & 116                  & 179            &  238             &  387               \\
$\mu_{\alpha}\cos \delta$ (mas yr$^{-1}$) & 2.503$\pm$0.004      & 0.257$\pm$0.010& -0.333$\pm$0.006& 1.006$\pm$0.007   \\
$\mu_{\delta}$ (mas yr$^{-1}$)            & -2.958$\pm$0.003     & -2.713$\pm$0.007 & -3.183 $\pm$0.005 & -2.584$\pm$0.005  \\
$d_{\varpi}$ (pc)                         & 1010$\pm$41          & 4114$\pm$500   &  2288$\pm$277    &  2674$\pm$457      \\
$d_{\rm BJ}$ (pc)                         & 973$\pm$37           & 3467$\pm$154   &  2111$\pm$74     &  2386$\pm$126      \\
$E(B-V)$ (mag)                            & $0.29^{+0.11}_{-0.09}$  & $0.52^{+0.22}_{-0.22}$  &  $0.93^{+0.33}_{-0.36}$   & $0.57^{+0.26}_{-0.27}$    \\
$A_{\rm V}$ (mag)                         & $0.91^{+0.36}_{-0.29}$  & $1.61^{+0.67}_{-0.67}$  &  $2.88^{+1.03}_{-1.11}$   & $1.78^{+0.80}_{-0.82}$    \\
$E({G_{\rm BP}-G_{\rm RP}})$ (mag)        & $0.41^{+0.16}_{-0.13}$  & $0.73^{+0.30}_{-0.30}$  &  $1.29^{+0.46}_{-0.50}$   & $0.80^{+0.36}_{-0.37}$    \\
$A_{\rm G}$ (mag)                         & $0.76^{+0.30}_{-0.24}$  & $1.35^{+0.56}_{-0.56}$  &  $2.41^{+0.86}_{-0.93}$   & $1.49^{+0.67}_{-0.69}$    \\
$[{\rm Fe/H}]$ (dex)                      & $-0.09^{+0.17}_{-0.14}$ & $-0.35^{+0.08}_{-0.14}$ &  $-0.16^{+0.08}_{-0.07}$  & $-0.24^{+0.19}_{-0.18}$   \\
Z                                         & $0.0124^{+0.0056}_{-0.0033}$ & $0.0070^{+0.0014}_{-0.0019}$  &  $0.0107^{+0.0153}_{-0.0107}$  &  $0.0087^{+0.0101}_{-0.0087}$     \\
$G - M_{\rm G}$ (mag)                     & $10.70^{+0.58}_{-0.45}$      & $14.05^{+0.71}_{-0.71}$        & $14.15^{+1.08}_{-1.27}$        & $13.47^{+0.81}_{-0.94}$    \\
$d_{\rm iso}$ (pc)                        & $972^{+136}_{-88}$           & $3472^{+242}_{-243}$    &  $2231^{+235}_{-326}$   & $2485^{+169}_{-269}$     \\
$\log t$ (yr)                             & $8.08^{+0.30}_{-0.49}$       & $9.40^{+0.36}_{-0.62}$  &  $8.36^{+0.51}_{-0.41}$ & $8.78^{+0.58}_{-0.49}$   \\
$X$ (pc)                                  & -965                 & -3435                & -2199                & -2444 \\
$Y$ (pc)                                  & -117                 & -486                 & -369                 & -417  \\
$Z$ (pc)                                  & 7                    & -135                 & 73                   & 169   \\
$R_{\rm gc}$ (pc)                         & $8966^{+135}_{-87}$  & $11445^{+241}_{-242}$& $10206^{+233}_{-323}$& $10452^{+167}_{-266}$ \\
PDMF slope                                & 1.16$\pm$0.35        & 1.62$\pm$0.47        & 1.44$\pm$0.32        & 1.61$\pm$0.18   \\
$T_{\rm E}$ (Myr)                         & $29.3^{+10.0}_{-6.0}$& $22.1^{+3.1}_{-2.4}$ & $49.6^{+9.7}_{-7.4}$ & $98.2^{+18.2}_{-13.3}$\\
$V_{\rm R}$ (km s$^{-1}$)                 & -4.99$\pm$3.28       & 15.57$\pm$5.29       & 39.30$\pm$0.31       & 39.80$\pm$2.75 \\
$U_{\rm LSR}$ (km s$^{-1}$)               & 15.92$\pm$3.02       & -0.45$\pm$4.93       & -25.91$\pm$0.31      & -25.03$\pm$2.24 \\
$V_{\rm LSR}$ (km s$^{-1}$)               & -2.60$\pm$2.39       & -26.38$\pm$3.30      & -19.90$\pm$3.52      & -24.77$\pm$3.31 \\
$W_{\rm LSR}$ (km s$^{-1}$)               & 9.89$\pm$0.74        & -11.66$\pm$1.03      & -11.21$\pm$2.31      & 5.81$\pm$0.21 \\
$S_{\rm LSR}$ (km s$^{-1}$)               & 18.92$\pm$3.92       & 28.84$\pm$6.02       & 34.54$\pm$4.23       & 35.70$\pm$4.01 \\
$Z_{\rm max}$ (pc)                        & 181$\pm$14           & 283$\pm$15           & 235$\pm$57           & 240$\pm$16 \\
$R_{\rm a}$ (pc)                          & 9569$\pm$60          & 11516$\pm$169        & 10410$\pm$235        & 10599$\pm$237 \\
$R_{\rm p}$ (pc)                          & 8266$\pm$73          & 9218$\pm$23          & 8553$\pm$23          & 8409$\pm$77 \\
$e$                                       & 0.07$\pm$0.01        & 0.11$\pm$0.01        & 0.09$\pm$0.01        & 0.12$\pm$0.03 \\
Birthplace (pc)                           & 8445$\pm$54          & 11302$\pm$733        & 9597$\pm$705         & 10549$\pm$577 \\
$P_{\rm orb}$ (Myr)                       & 252$\pm$1            & 299$\pm$3            & 270$\pm$5            & 271$\pm$3 \\
$R_{\rm teo}$ (pc)                        & 8866                 & 10238                & 9388                 & 9379 \\
\hline
        \end{tabular}%
    \label{tab:Final_table}%
    
\end{table*}%


\section{Discussion and Conclusion}\label{sec:discuss}

This study presents a detailed investigation of the OCs COIN-Gaia~24, Czernik~24, FSR~0893, and UBC~74, which are located toward the Galactic anti-center. This region holds strategic significance because its position opposite the Galactic center offers an unobstructed view of the outer disc, where the dynamical influence of the Galactic bar is minimal. Consequently, it provides an excellent laboratory for studying large-scale Galactic features such as the outer disc structure, metallicity gradients, disc flaring, and warping. Despite their location in a key region, these OCs have received limited attention in the literature, which motivates their selection for the present detailed investigation.

We conducted a comprehensive examination of the structural, astrophysical, kinematic, and dynamical characteristics of these four OCs, based on data from the {\it Gaia} DR3 catalog \citep{Gaia23}, in this study. The astrophysical and kinematic parameters, as well as the Galactic orbital properties, were derived using stars with membership probabilities of $P \geq 0.5$, identified through the \texttt{UPMASK} algorithm \citep{Krone-Martins14}. The principal results of the study can be summarized as follows:

\begin{itemize}
\item Membership selection using the \texttt{UPMASK} algorithm in a five-dimensional parameter space identified 116 stars in COIN-Gaia~24, 179 stars in Czernik~24, 238 stars in FSR~0893, and 387 stars in UBC~74 as most probable cluster members ($P \ge 0.5$) within the photometric completeness limit ($G \le 20.5$ mag).

\item The radial density profiles (RDPs) of the clusters were fitted with the \citet{King62} model to derive their structural parameters. The resulting core radii ($r_{\rm c}$) are $1.91^{+0.34}_{-0.27}$ pc, $1.38^{+0.17}_{-0.15}$ pc, $2.77^{+0.34}_{-0.32}$ pc, and $5.37^{+0.77}_{-0.67}$ pc for COIN-Gaia~24, Czernik~24, FSR~0893, and UBC~74, respectively. The corresponding concentration parameters ($C$) were estimated as $9.26^{+0.62}_{-0.01}$ for COIN-Gaia~24, $10.26^{+0.08}_{-0.43}$ for Czernik~24, $6.84^{+0.05}_{-0.21}$ for FSR~0893, and $3.82^{+0.16}_{-0.29}$ for UBC~74. These values indicate moderate to high central concentration for Czernik 24 and COIN-Gaia~24, while UBC~74 appears to be a relatively loose and extended system.

\item The mean proper-motion components $\langle\mu_{\alpha}\cos\delta,\mu_{\delta}\rangle$ were derived as $(2.503\pm 0.004,-2.958\pm 0.003)$ mas yr$^{-1}$ for COIN-Gaia~24, $(0.257\pm 0.010,-2.713\pm 0.007)$ mas yr$^{-1}$ for Czernik~24, $(-0.333\pm 0.006,-3.183\pm 0.005)$ mas yr$^{-1}$ for FSR~0893, and $(1.006\pm 0.007,-2.584\pm 0.005)$ mas yr$^{-1}$ for UBC~74. Distances derived from trigonometric parallaxes ($d_{\varpi}$) were found to be $1010\pm 41$, $4114\pm 500$, $2288\pm 277$, and $2674\pm 457$ pc, respectively.

\item Reddening and metallicity were determined using two methods: MCMC isochrone fitting and SED analysis. The adopted MCMC results for metallicity ([Fe/H]) are $-0.09^{+0.17}_{-0.14}$ dex for COIN-Gaia~24, $-0.35^{+0.08}_{-0.14}$ dex for Czernik~24, $-0.16^{+0.08}_{-0.07}$ dex for FSR~0893, and $-0.24^{+0.19}_{-0.18}$ dex for UBC~74. The corresponding $G$-band extinctions ($A_{\rm G}$) are $0.76^{+0.30}_{-0.24}$, $1.35^{+0.56}_{-0.56}$, $2.41^{+0.86}_{-0.93}$, and $1.49^{+0.67}_{-0.69}$ mag., respectively.

\item Distances and ages of the OCs were determined using MCMC analyses with \texttt{PARSEC} isochrones. The isochrone distances ($d_{\rm iso}$) were found to be $972^{+136}_{-88}$ pc for COIN-Gaia~24, $3472^{+242}_{-243}$ pc for Czernik~24, $2231^{+235}_{-326}$ pc for FSR~0893, and $2485^{+169}_{-269}$ pc for UBC~74. The ages ($\log t$) were determined as $8.08^{+0.30}_{-0.49}$, $9.40^{+0.36}_{-0.62}$, $8.36^{+0.51}_{-0.41}$, and $8.78^{+0.58}_{-0.49}$, respectively. These values show consistency with parameters derived from the independent SED analysis.

\item The mean radial velocities ($\langle V_{\rm R}\rangle$) were calculated from {\it Gaia} DR3 member stars. The values were determined to be $-4.99\pm 3.28$ km s$^{-1}$ (from 19 stars) for COIN-Gaia 24, $15.57\pm 5.29$ km s$^{-1}$ (from 11 stars) for Czernik 24, $39.30\pm 0.31$ km s$^{-1}$ (from 7 stars) for FSR 0893, and $39.80\pm 2.75$ km s$^{-1}$ (from 9 stars) for UBC 74.

\item The Galactic orbital analyses, performed using \texttt{galpy}, revealed that all four OCs originated outside the solar circle and have remained confined to that region. The derived orbital parameters, including low eccentricities ($e\leq0.12$) and maximum vertical distances from the Galactic plane ($Z_{\rm max} \le 283$ pc), corroborate their classification as members of the Galactic thin-disc population.

\item MS stars were selected from the turn-off down to fainter magnitudes in each cluster. The corresponding mass ranges are $0.55<M/M_{\odot}\leq3.1$ for COIN-Gaia~24, $0.9< M/M_{\odot}\leq1.35$ for Czernik~24, $0.7< M/M_{\odot}\leq1.7$ for FSR~0893, and $0.8< M/M_{\odot}\leq1.9$ for UBC~74. The high completeness of {\it Gaia} photometry ensures robust PDMF slopes, thereby minimizing biases in low-mass stars. 

\item The dynamical relaxation times ($T_{\rm E}$) for COIN-Gaia 24, Czernik 24, FSR 0893, and UBC 74 were calculated as $29.3^{+10.0}_{-6.0}$ Myr, $22.1^{+3.1}_{-2.4}$ Myr, $49.6^{+9.7}_{-7.4}$ Myr, and $98.2^{+18.2}_{-13.3}$ Myr, respectively. Comparing these with the OC ages, the ratios $\tau_{\text{age}}/T_{\rm E}$ were found to be  $4.1$, $113.7$, $4.6$, and $6.1$. These results suggest the clusters are in a state of advanced dynamical relaxation, although these $T_{\rm E}$ values represent formal lower limits based on the observed stellar census. 
\end{itemize}

The astrometric parameters derived in this study for the four OCs show strong consistency with recent \textit{Gaia}~DR3-based catalogs. For COIN-Gaia~24, the mean proper motions of $\mu_{\alpha}\cos\delta = 2.503 \pm 0.004~\mathrm{mas~yr^{-1}}$ and $\mu_{\delta} = -2.958 \pm 0.003~\mathrm{mas~yr^{-1}}$, along with a distance of $1010 \pm 41~\mathrm{pc}$, are in good agreement with the values reported by \citet{Liu19} and \citet{Cantat-Gaudin20}. Czernik~24 exhibits $\mu_{\alpha}\cos\delta = 0.257 \pm 0.010~\mathrm{mas~yr^{-1}}$, $\mu_{\delta} = -2.713 \pm 0.007~\mathrm{mas~yr^{-1}}$, and $d = 4114 \pm 500~\mathrm{pc}$, consistent with the 3.5-4~kpc range and similar proper-motion components listed in recent catalogs \citep{Dias21, Hunt2024}. For FSR~0893, the proper motions $\mu_{\alpha}\cos\delta = -0.333 \pm 0.006~\mathrm{mas~yr^{-1}}$ and $\mu_{\delta} = -3.183 \pm 0.005~\mathrm{mas~yr^{-1}}$, together with a distance of $2288 \pm 277~\mathrm{pc}$, closely match the latest $Gaia$ DR3-based determinations by \citet{Cantat-Gaudin20} and \citet{Poggio2021}. Similarly, UBC~74 shows $\mu_{\alpha}\cos\delta = 1.006 \pm 0.007~\mathrm{mas~yr^{-1}}$, $\mu_{\delta} = -2.584 \pm 0.005~\mathrm{mas~yr^{-1}}$, and $d = 2674 \pm 457~\mathrm{pc}$, in strong agreement with the results of \citet{Cantat-Gaudin_Anders20} and \citet{Tarricq2022}. Overall, these findings confirm the robustness of the membership selection and astrometric calibration procedures adopted in this study.

A comparison between our newly derived astrophysical parameters and those reported in Table~\ref{tab:literature} shows overall consistency, with minor discrepancies likely arising from differences in data quality and analysis methods. For COIN-Gaia~24, the derived age of $8.08^{+0.30}_{-0.49}$ agrees well with the previously reported 60–300~Myr range. Its distance ($d_{\rm iso}=972^{+136}_{-88}$~pc) and colour excess ($E({G_{\rm BP}-G_{\rm RP}})=0.41^{+0.16}_{-0.13}$~mag) are consistent with earlier estimates of 900-1150~pc and $E(B-V)\sim0.2-0.4$~mag. For Czernik~24, our results of $9.40^{+0.36}_{-0.62}$ (about 2.5~Gyr) and $d_{\rm iso}=3472^{+242}_{-243}$~pc lie near the midpoint of the wide literature ranges (1.2-2.7~Gyr and 3.4-4~kpc), while the colour excess $E({G_{\rm BP}-G_{\rm RP}})=0.73^{+0.30}_{-0.30}$~mag is in good agreement with the high reddening values (0.5-0.75~mag) previously reported. FSR~0893 shows $8.36^{+0.51}_{-0.41}$ and $d_{\rm iso}=2231^{+235}_{-326}$~pc, in line with the 1.8-2.7~kpc distance and 200-1000~Myr age estimates from earlier studies; its metallicity, $[{\rm Fe/H}]=-0.16^{+0.08}_{-0.07}$~dex, also agrees well with the literature median of about $-0.10$~dex. For UBC~74, we find $8.78^{+0.58}_{-0.49}$ ($\approx$ 600~Myr) and $d_{\rm iso}=2485^{+169}_{-269}$~pc, consistent with the 2.4-2.8~kpc range reported in previous works, while its metallicity of $[{\rm Fe/H}]=-0.24^{+0.19}_{-0.18}$~dex confirms its slightly metal-poor nature. Overall, this study demonstrates strong consistency with previous photometric and kinematic results, while significantly reducing parameter uncertainties due to the high precision of the \textit{Gaia} EDR3 data.

The application of two independent and robust approaches, MCMC-based isochrone fitting and stellar SED analysis, for determining astrophysical parameters, together with the strong agreement between the derived distance, extinction, and metallicity values, reinforces the reliability of our results. The investigated OC sample spans a wide range of ages: COIN-Gaia 24 is a young system (log $t$ $\approx$ 8.08), Czernik 24 is an old cluster (log $t$ $\approx$ 9.40), while FSR 0893 and UBC 74 represent intermediate-age populations (log $t$ $\approx$ 8.36-8.78). All four clusters exhibit slightly subsolar metallicities ([Fe/H] = -0.09 to -0.35 dex), consistent with the negative radial metallicity gradient typically observed in the outer Galactic disc. In particular, Czernik 24, the oldest and most distant cluster in the sample (log $t$ $\approx$ 9.40, $d$ $\approx$ 3.5 kpc), also shows the lowest metallicity ([Fe/H] $\approx$ -0.35 dex), further supporting this trend.

The Galactic disc positions of the OCs were investigated through kinematic and orbital analyses. Orbital integrations based on mean radial velocities and astrometric parameters from {\it Gaia} DR3 reveal that all four clusters are located beyond the solar circle ($R_{\rm gc}$ $\approx$ 9-11.5 kpc) and have remained in this region throughout their orbital evolution. The derived low eccentricities ($e\leq 0.12$) and small maximum vertical excursions from the Galactic plane ($Z_{\rm max}$ $\leq$ 283 pc) kinematically confirm that these clusters belong to the thin disc population \citep{Plevne15}. 

The four OCs exhibit moderate vertical amplitudes in the axisymmetric potential (MW14), with $Z_{\max} \approx 180-280$ pc and eccentricities $e \approx 0.07$--$0.12$, consistent with mildly heated thin-disc orbits spanning $\langle R_{\rm gc} \rangle \approx 9-11.5$ kpc. When evaluated relative to the warped midplane, the effective vertical amplitudes rise to $Z_{\rm max, wf} \approx 0.27-0.55$ kpc, corresponding to an increase of roughly 50--95\% (see Table~\ref{tab:tab:warp-flare}). This mainly reflects the downward offset of the warped midplane at the relevant azimuths, whereas the cluster orbits themselves remain approximately symmetric about the geometrical plane $Z \approx 0$ pc. The warp matching fractions are generally low, ranging from $\sim 26-43\%$, and typical scale-height-normalized elevations remain modest, with median $(|Z|/h_{\rm z}) \approx 0.4-0.6$. At their present locations, the OCs occupy heights of $|Z_{\rm now}| \approx (0.1-0.6)h_{\rm z}$, while the local flare increases the thin-disc scale height by only $\sim 5$--$17\%$.

Overall, these findings indicate that the clusters do not closely follow either the Galactic warp or the outer-disc flare. Instead, they are consistent with typical thin-disc members whose orbits pass through a mildly warped and slightly flared outer disc. The geometric warp correction affects only the inferred vertical amplitudes, leaving the orbital eccentricities and mean radii unchanged within numerical precision. Therefore, the main conclusions regarding the clusters' radial migration histories and present-day kinematics are not significantly impacted by neglecting warp or flare in the adopted axisymmetric potential.

The principal strength of this study lies in the cross-validation of key astrophysical parameters (extinction, distance, metallicity, and age) derived from two independent Bayesian approaches: an MCMC-based isochrone fitting and a multi-wavelength SED analysis employing the \texttt{ARIADNE} framework, which combines data from {\it Gaia}, 2MASS, and {\it WISE}. This dual approach allowed for a precise characterization of differential reddening effects, particularly evident in FSR 0893. The accuracy of the atmospheric model parameters ($T_{\rm eff}$, $\log g$, [Fe/H]) derived in this study was confirmed through a direct comparison with the {\it Gaia} DR3 XP spectroscopic catalog \citep{Kordopatis2024}, revealing excellent agreement. Building upon this solid parameter foundation, our kinematic analysis, employing \texttt{galpy} for precise orbital integrations, conclusively establishes that all four OCs are highly probable members of the Galactic thin disc. The identification of BSSs in Czernik~24 and variable star candidates in COIN-Gaia~24 further enriches the astrophysical characterization of these OCs and adds new insights into the stellar populations of the outer Galactic disc.

The derived PDMF slopes for the analyzed OCs, $\Gamma=1.16\pm 0.35$ for COIN-Gaia~24, $\Gamma=1.62\pm 0.47$ for Czernik~24, $\Gamma=1.44\pm 0.32$ for FSR~0893, and $\Gamma=1.61\pm 0.18$ for UBC~74, can be compared against the canonical Salpeter value of $\Gamma = 1.35$ \citep{Salpeter55}. COIN-Gaia~24 exhibits a slightly flatter slope, indicating a relative overabundance of higher-mass stars compared to the Salpeter expectation, whereas Czernik~24 and UBC~74 show somewhat steeper slopes, suggesting a modest excess of low-mass stars.  The PDMF slope of FSR~0893 is very close to the Salpeter value, consistent with a standard mass distribution. Overall, the PDMF slopes of these OCs fall within the typical uncertainties expected for OCs \citep{Bukowiecki2012, Bisht19}, reflecting that while minor deviations from Salpeter are present, the global mass distributions are broadly consistent with canonical stellar formation theories \citep{Kroupa01, Hopkins2012}.

\begin{figure*}
\centering
\includegraphics[width=0.65\linewidth]{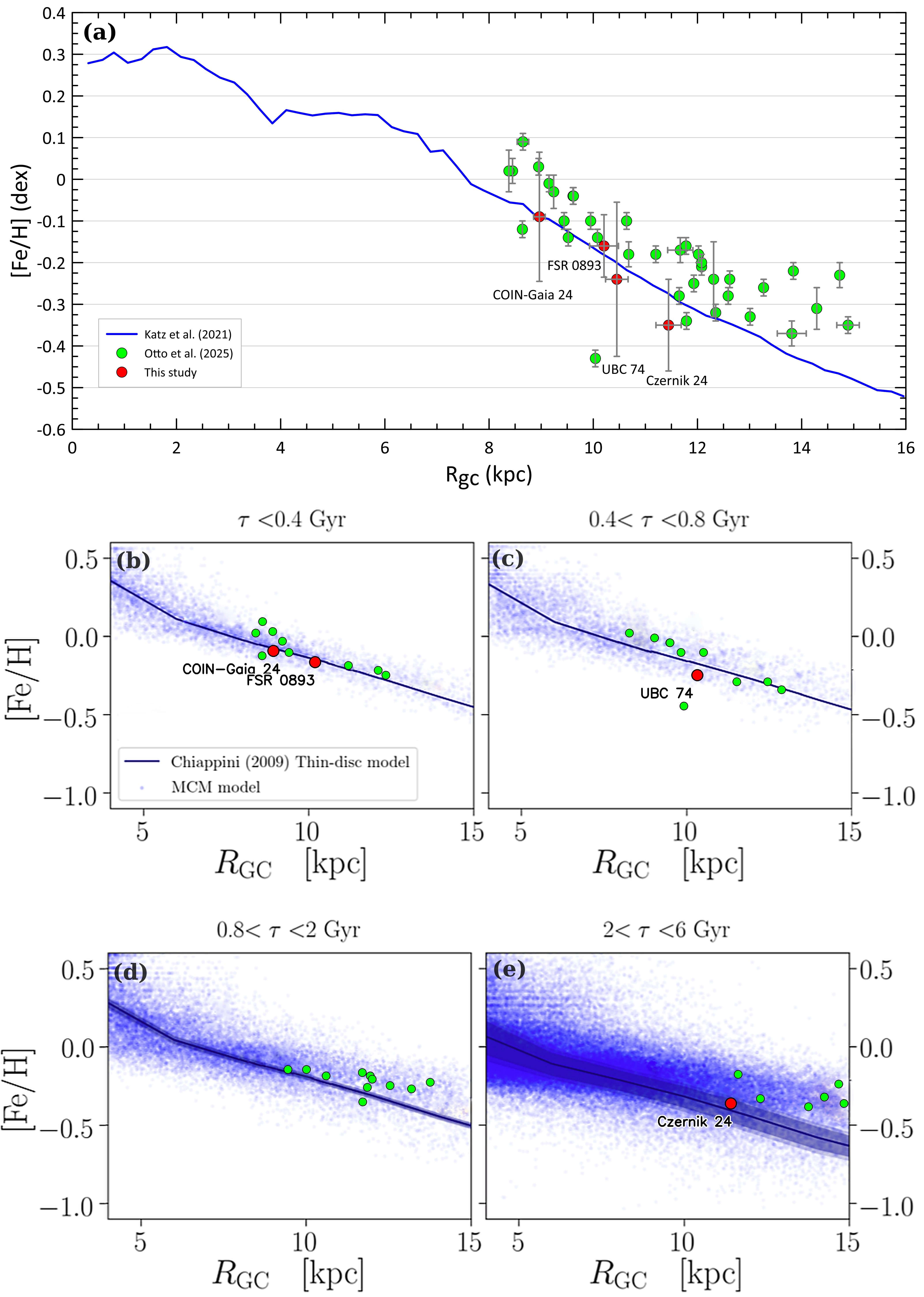}
\caption{Radial metallicity profiles of the Milky Way are presented in panel (a) for giant stars \citep[light blue curve;][]{Katz2021}, and in panels (b-e) for theoretical and simulation models \citep[blue lines and dots;][]{Chiappini2009, Minchev2013, Minchev2014}. The red circles indicate the four OCs analyzed in this study, and the green circles represent the OCs from \citet{Otto2025}.}
\label{fig: radial-metallicity}
\end{figure*}

We have found that the ratios $\tau_{\rm age} / T_E$ are approximately $4.1$, $113.7$, $4.6$, and $6.1$, for COIN-Gaia~24, Czernik~24, FSR~0893, and UBC~74, respectively. These values indicate that all four OCs have evolved well beyond their dynamical relaxation times, highlighting a mature dynamical state characterized by significant mass segregation and internal energy redistribution \citep[e.g.,][]{Alfonso2024, Angelo2025, Coenda2025}. In particular, the exceptionally high ratio for Czernik~24 underscores its highly evolved nature, suggesting an advanced degree of dynamical processing that may have strongly influenced its current stellar structure and kinematic properties \citep[see also,][]{Guszejnov2022}. Nevertheless, the calculated $T_{\rm E}$ values should be regarded as formal lower limits, as our estimates consider only the stellar population above the photometric completeness threshold. The presence of unresolved low-mass stars or binary systems would increase both the total number of members and the cluster mass, potentially extending the true relaxation timescales \citep[e.g.,][]{Bisht19, Bisht2026}. 

The location of the four OCs in the direction opposite to the Galactic center allows the metallicity profile of the Galactic disc to be examined using independent tracers. \citet{Katz2021} selected giant stars belonging to the thin-disc population from the APOGEE DR16 database \citep{Ahumada2020} and analyzed their [Fe/H] abundances, obtained from the APOGEE spectroscopic survey, as a function of Galactocentric distance. In this way, they investigated the radial variation of metallicity across the Galactic disc (Figure~\ref{fig: radial-metallicity}a). The slope of the metallicity gradient derived by \citet{Katz2021} is approximately $-0.07$ dex kpc$^{-1}$ over the distance range $6 < R_{\rm gc}~{\rm (kpc)} \leq 14$, which is consistent with values reported in previous studies and confirms the reliability of the APOGEE data in tracing the chemical evolution of the Galactic disc. In the present study, the four OCs, whose metallicities and distances were derived using the MCMC method, were plotted onto the plane defined by \citet{Katz2021}, and their metallicities were compared with the radial metallicity trend inferred from giant stars. Additionally, the mean metallicities of the OCs studied within the scope of the Open Cluster Chemical Abundances and Mapping \citep[OCCAM,][]{Frinchaboy2013} project were also marked on the same metallicity profile. Since the four OCs analyzed in this study were selected from the direction opposite to the Galactic center, they are comparable to the OCs in the sample of \citet{Otto2025}, from which 37 OCs with Galactic longitudes between $160^{\rm o}$ and $200^{\rm o}$ were identified. Investigation of the OCs studied by \citet{Otto2025} on the same profile shows that they occupy the metal-rich region relative to the metallicity curve obtained by \citet{Katz2021}. A combined evaluation of both studies reveals that the metallicity gradient remains clearly defined up to approximately 12 kpc, beyond which a distinct ``knee'' structure emerges \citep[e.g.][]{Carraro2007, Villanova2007, Cunha2016, Netopil22}. The results show that the radial metallicity profile derived from giant stars is consistent, within observational uncertainties, with the metallicities of the four investigated OCs.

Furthermore, the four OCs were tested against the chemical evolution model of the Galactic thin disc proposed by \citet{Chiappini2009}. In the analysis, the inclusion of additional OCs selected from the sample of \citet{Otto2025} into the panels containing the chemical evolution models increased the number of clusters available for comparison. When the ages of the OCs used in this study, together with those from \citet{Otto2025}, are taken into account and placed on the corresponding panels of the chemical evolution model, their distributions exhibit a strong overall agreement with Chiappini’s two-infall scenario \citep{Chiappini2009}. However, in the panel representing the oldest populations, while the Czernik 24  examined in this study aligns well with the model predictions, the clusters from \citet{Otto2025} occupy regions that are more metal-rich than the model curve. One possible reason for the higher metallicities of these clusters is the small number of stars sampled within them (fewer than four), which may increase the uncertainties in the metallicity estimates. Another possibility is that OCs located in the outer regions of the Galactic disc are statistically older, making them more susceptible to Galactic gravitational effects, which in turn may influence their Galactic orbits. Overall, the consistency between the chemical evolution models and the distributions of the investigated OCs suggests that the outer Galactic disc has experienced a chemical enrichment history compatible with inside-out disc formation scenarios, reinforcing the view that OCs are reliable tracers of the chemical and structural evolution of the Galactic disc \citep[c.f.][]{Frinchaboy2013, Myers2022, Netopil22, Otto2025}.

\section*{Acknowledgements}

 This study has been supported in part by the Scientific and Technological Research Council (T\"UB\.ITAK) 122F109. We also made use of SIMBAD and VizieR databases at CDS, Strasbourg, France. This research has made use of the Astrophysics Data System, funded by NASA under Cooperative Agreement 80NSSC21M0056. We made use of data from the European Space Agency (ESA) mission \emph{Gaia}\footnote{https://www.cosmos.esa.int/gaia}, processed by the \emph{Gaia} Data Processing and Analysis Consortium (DPAC)\footnote{https://www.cosmos.esa.int/web/gaia/dpac/consortium}. Funding for DPAC has been provided by national institutions, in particular the institutions participating in the \emph{Gaia} Multilateral Agreement. 



\bibliographystyle{raa}
\bibliography{open_clusters}

\end{document}